\documentclass[a4paper,12pt]{article}

\usepackage[T1]{fontenc}
\usepackage{amsmath,amssymb,amsthm,mathtools}
\usepackage{graphicx}
\usepackage{enumitem}
\usepackage{natbib}
\usepackage{tikz}
\usetikzlibrary{shapes}
\usepackage{chngcntr}

\allowdisplaybreaks

\usepackage{authblk}

\renewenvironment{abstract}{%
	\vspace{6pt}%
	\begin{center}%
		\begin{minipage}{320pt}%
			\small%
			\begin{center}%
				\textbf{Abstract}%
			\end{center}%
		}{\end{minipage}\end{center}}
\newcommand{\keywords}[1]{%
	\begin{center}%
		\begin{minipage}{320pt}%
			\textbf{Keywords:}~{#1}
		\end{minipage}%
	\end{center}%
}

\begin{document}
\title{Modelling columnarity of pyramidal cells in the human cerebral cortex}
\author{Andreas Dyreborg Christoffersen, Jesper M\o ller and Heidi S\o gaard Christensen}
\affil{Aalborg University}

\maketitle

\begin{abstract}
For modelling the location of pyramidal cells in the human cerebral cortex we suggest a hierarchical point process in $\mathbb{R}^3$ that exhibits anisotropy in the form of cylinders extending along the $z$-axis.
The model consists first of a generalised shot noise Cox process for the $xy$-coordinates, providing cylindrical clusters, and next of a Markov random field model for the $z$-coordinates conditioned on the $xy$-coordinates, providing either repulsion, aggregation, or both within specified areas of interaction. Several cases of these hierarchical point processes are fitted to two pyramidal cell datasets, and of these a final model allowing for both repulsion and attraction between the points seem adequate. We discuss how the final model relates to the so-called minicolumn hypothesis in neuroscience.
\end{abstract}


\section{Introduction}
The structuring of neurons in the human brain is a subject of great interest since abnormal structures may be linked to certain neurological diseases \citep[see][]{Buxhoeveden:Casanova:2002,Casanova:2007,Casanova:etal:2006,Esiri:Chance:2006,Chance:etal:2011}. A specific structure that has been extensively studied in the biological literature is the so called 'minicolumn' structure of the cells in the cerebral cortex \citep[see][and references therein]{Buxhoeveden:Casanova:2002,Rafati:etal:2016}. \cite{Rafati:etal:2016} characterised these minicolumns as `linear aggregates of neurons organised vertically in units that traverse the cortical layer II--VI, and have in humans a diameter of 35--60\,$\mu\mathrm{m}$ and consist typically of 80--100 neurons'.

\subsection{Data}
In this paper we analyse the structuring of pyramidal cells, which make up approximately 75--80\% of all neurons \citep{Buxhoeveden:Casanova:2002} and are pyramid shaped cells, where the so-called apical dendrite extends from the top/apex of a pyramidal cell. 
Specifically, the paper is concerned with pyramidal cells from the so-called Brodmann area four of the human cerebral cortex. The neocortex constitutes most of the cerebral cortex and can be divided into six layers. We consider two datasets consisting of the locations and orientations of pyramidal cells in a section of the third and fifth layer, respectively.
Here, each location is a three-dimensional coordinate representing the centre of a pyramidal cell's nucleolus and each orientation is a unit vector representing the apical dendrite's position relative to the corresponding nucleolus.

\begin{figure}[!ht]
	\centering
	\includegraphics[scale=.4, trim={26mm 30mm 16mm 25mm}, clip]{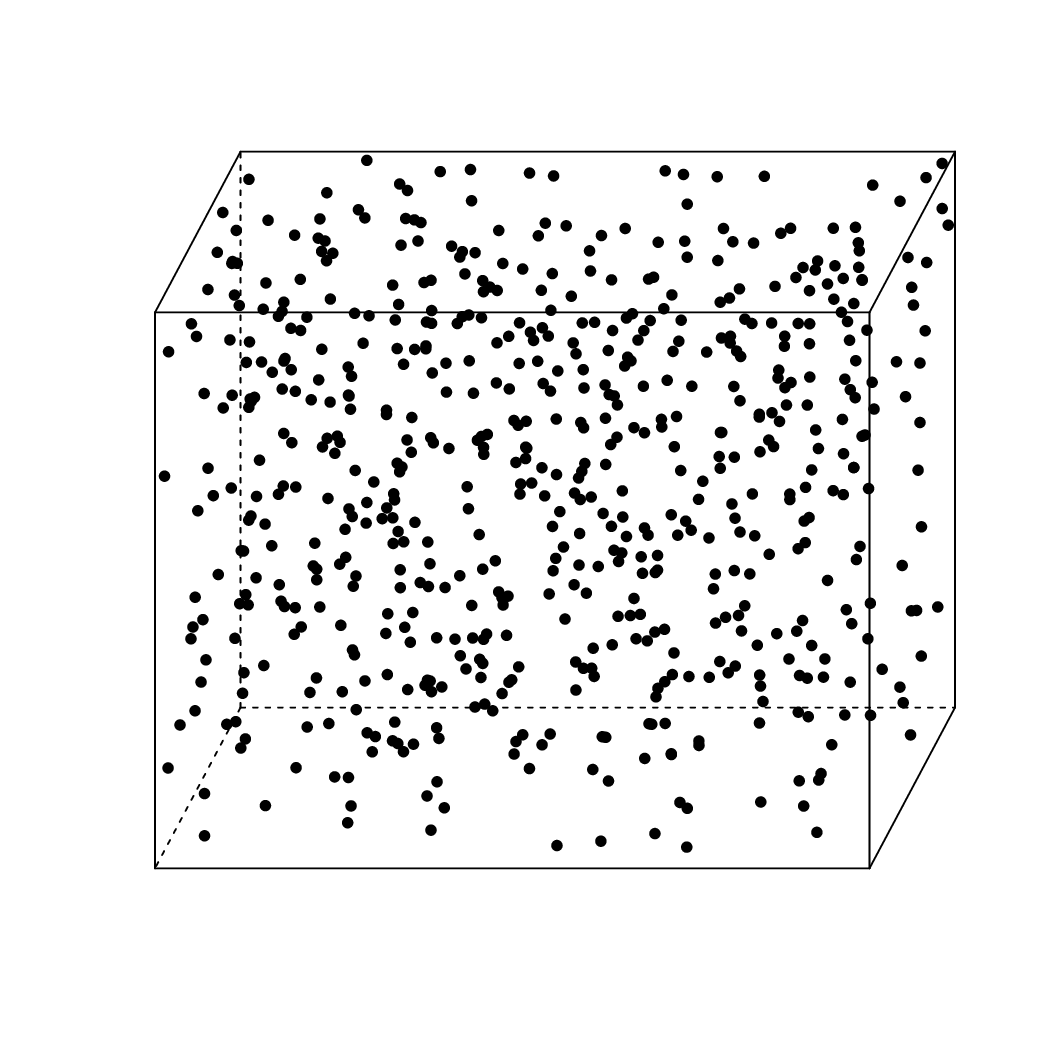}
	\hfill 
	\includegraphics[scale=.4, trim={2mm 25mm 2mm 62mm}, clip]{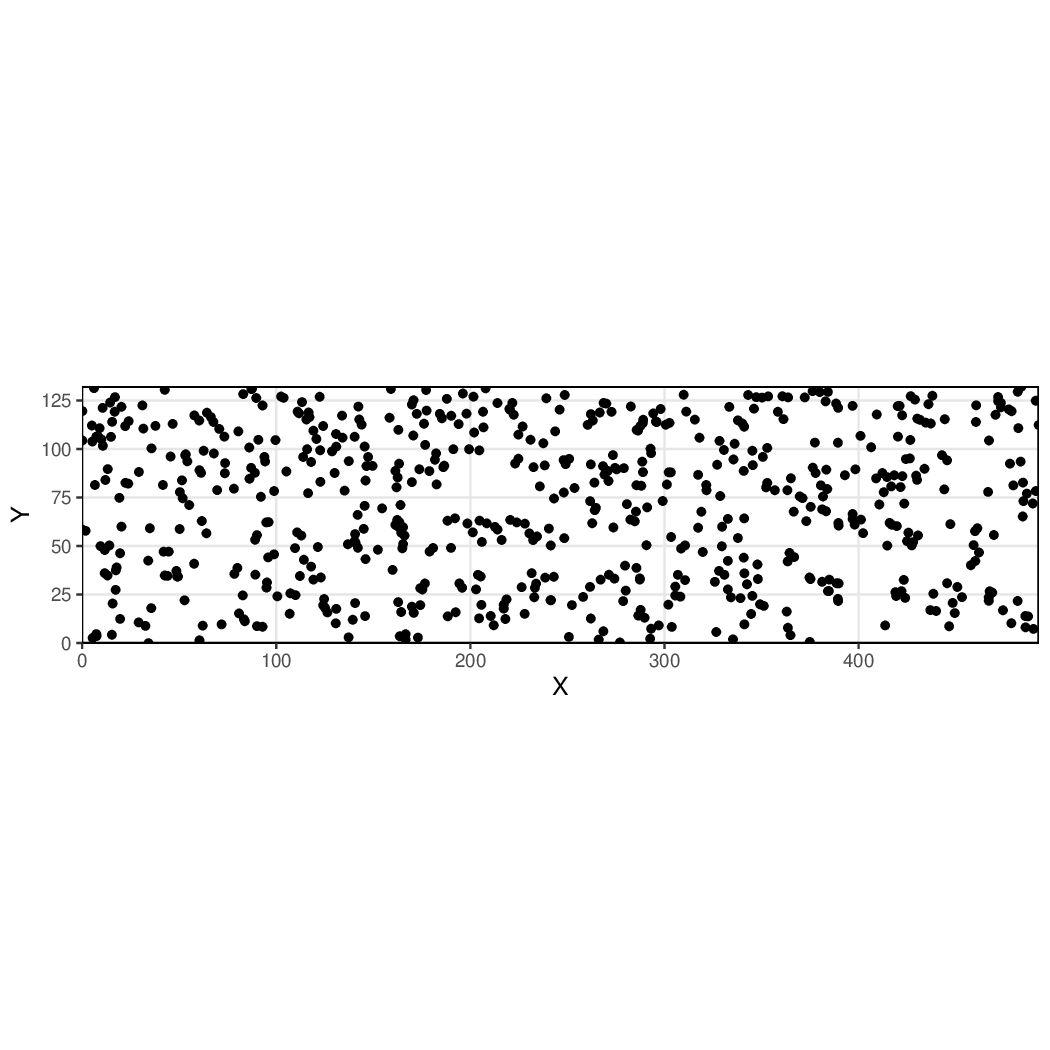}\\
	\includegraphics[scale=.4, trim={26mm 30mm 16mm 25mm}, clip]{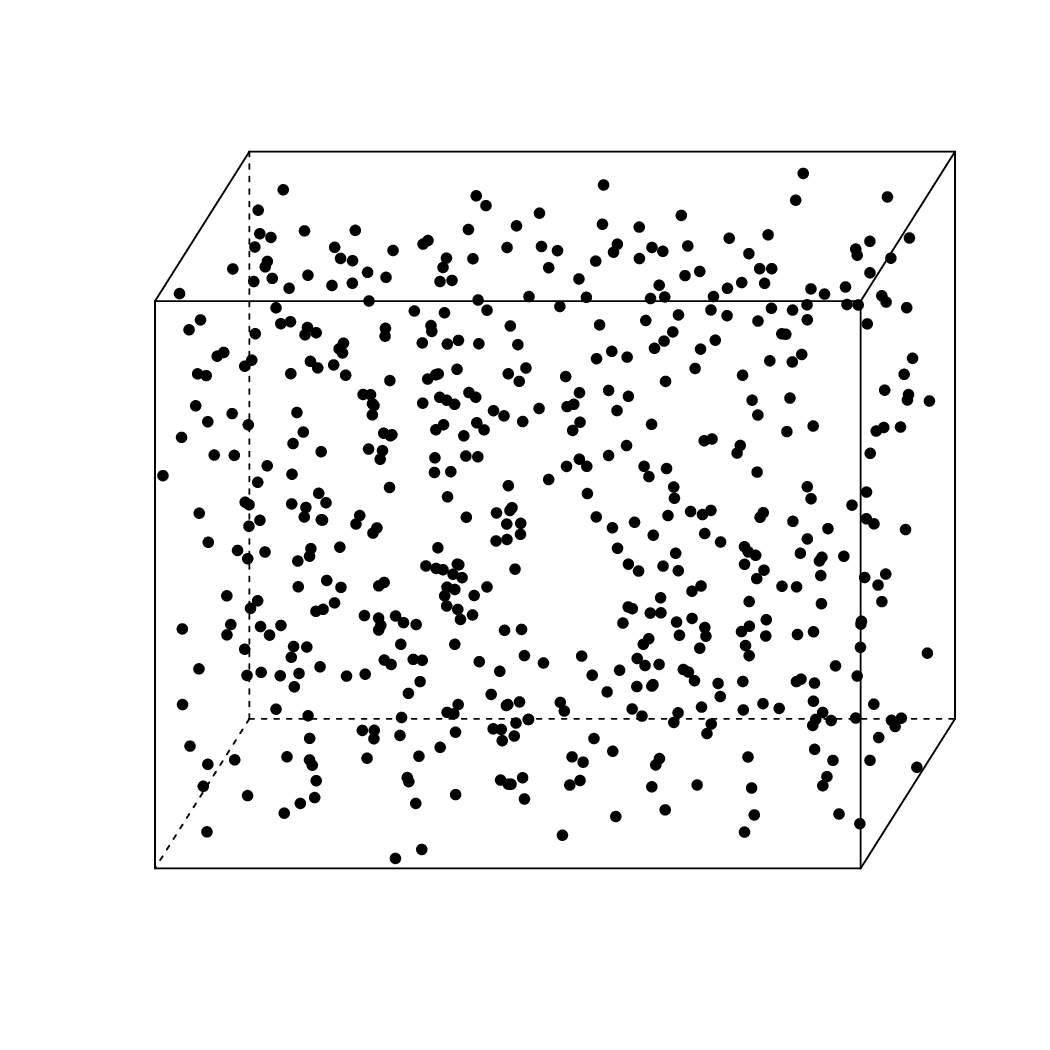}
	\hfill
	\includegraphics[scale=.4, trim={2mm 25mm 2mm 62mm}, clip]{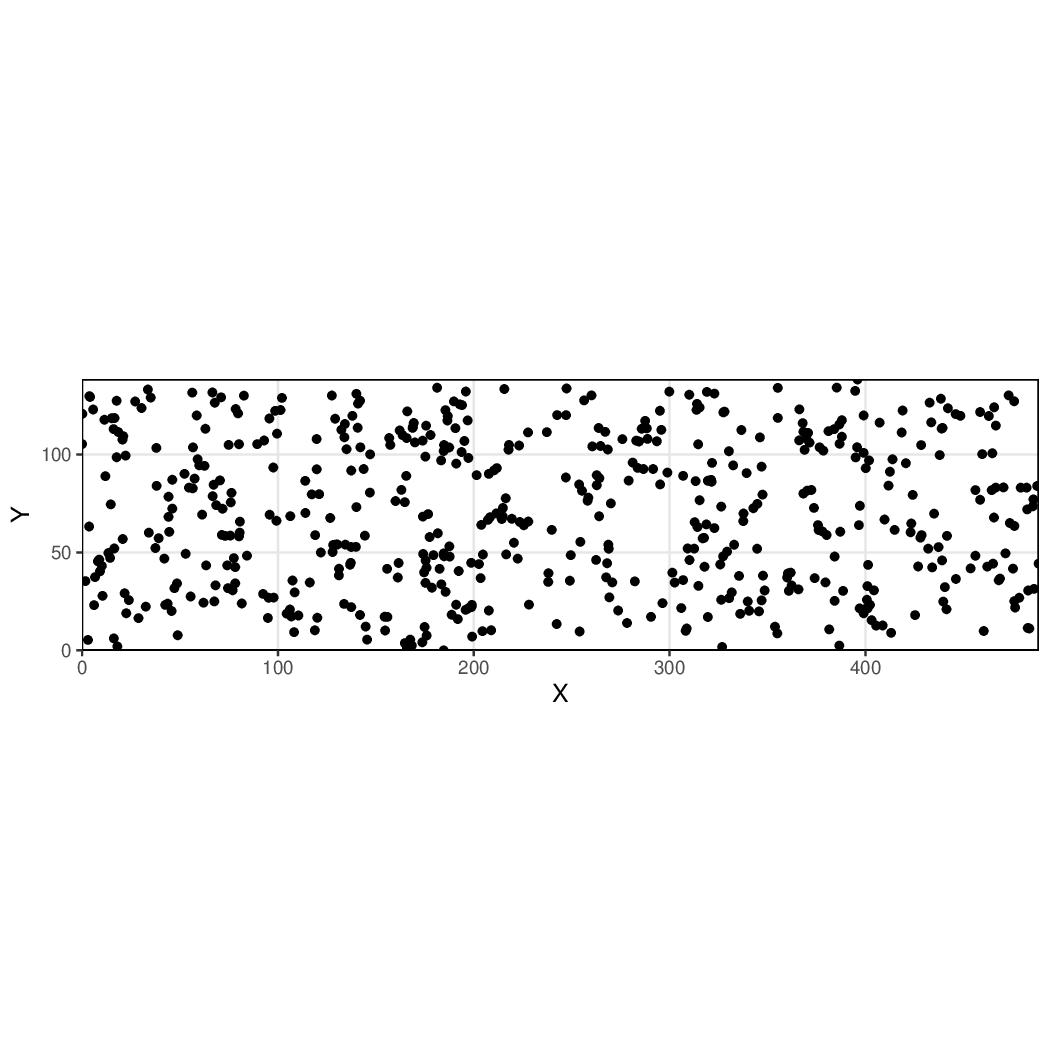}
	\caption{
		3D point patterns (left) and projections onto the $xy$-plane (right) of the nucleolus locations for datasets \texttt{L3} (top) and \texttt{L5} (bottom).
	}
	\label{f:data}
\end{figure}

The left part of Figure~\ref{f:data} shows two point pattern datasets of 634 and 548 nucleolus locations which will be referred to as \texttt{L3} {(top)} and \texttt{L5} {(bottom)}, respectively \citep[for plot of the orientations for \texttt{L3}, see][]{Moller:etal:2019}. Note that the observation window $W$ for the cell locations is a rectangular region with side lengths 492.70\,$\mu\mathrm{m}$, 132.03\,$\mu\mathrm{m}$, and 407.70\,$\mu\mathrm{m}$ for \texttt{L3} and 488.40\,$\mu\mathrm{m}$, 138.33\,$\mu\mathrm{m}$, and 495.40\,$\mu\mathrm{m}$ for \texttt{L5}.
Notice also that the nucleolus locations are recorded such that the $z$-axis is perpendicular to the so-called pial surface of the brain. In accordance to the minicolumn hypothesis, this implies that the minicolumns extend parallel to the $z$-axis. In the right part of Figure~\ref{f:data} we have therefore shown the 2D $xy$-locations of the two 3D point pattern datasets. 

\subsection{Background and objective}
\cite{Moller:etal:2019} found independence between locations and orientations for \texttt{L3} meaning that the two components may be modelled separately; the same conclusion has afterwards been drawn for \texttt{L5}. As they also found a suitable inhomogeneous Poisson process model for the orientations, and since it is {difficult to visually detect any} structure in the point patterns shown in Figure~\ref{f:data}, the focus of this paper is on modelling the nucleolus locations. 
In particular, we aim at modelling the nucleolus locations for \texttt{L3} respective \texttt{L5} by a spatial point process with a columnar structure and discuss to what extent this relates to the minicolumn hypothesis. Note that for the two datasets we use the same notation $X$ for the spatial point process, and we view $X$ as a random finite subset of $W$.  

To the best of our knowledge the so-called Poisson line cluster point process \citep[see][]{Moller:etal:2016} is the only existing point process model for modelling columnar structures. This model was considered by \cite{Rafati:etal:2016} in connection to another dataset of pyramidal nucleolus locations, but it was not fitted to that dataset.
For each point pattern considered in the present paper, we notice later that a more advanced model than the Poisson line cluster point process is needed; below we describe such a model for $X$.

\subsection{Hierarchical point process models}\label{s:models}
We consider a hierarchical model for $X$ as follows. Note that the observation window is a product space, $W=W_{xy}\times W_z$, where $W_{xy}$ is a rectangular region in the $xy$-plane and $W_z$ is an interval on the $z$-axis. Let $X_{xy}$ be the point process of $xy$-coordinates in $X$ (i.e.\ given by the projection of $X$ onto the $xy$-plane), associate to each point $(x_i,y_i)\in X_{xy}$ the corresponding $z$-coordinate $z_i$ so that $(x_i,y_i,z_i)\in X$, and conditioned on the $xy$-points $X_{xy}=\{(x_i,y_i)\}_{i=1}^n$, let $X_z=(z_i)_{i=1}^n$ be the random vector of $z$-coordinates in $X$ (with an arbitrary ordering of these $n$ points and where $(z_i)_{i=1}^n$ is short hand notation for $(z_1,\ldots,z_n)$). Clearly, $X$ is in a one-to-one correspondence with $X_{xy}$ and $X_z$, and we model first $X_{xy}$ and second $X_z$ conditioned on $X_{xy}$. Further details are given below and in Sections~\ref{s:CSR}--\ref{s:gen_dPLCPP}.

\subsubsection{The model for $X_{xy}$}\label{s:model1}
For $X_{xy}$ we consider the restriction of a cluster point process in $\mathbb R^2$ to $W_{xy}$ defined as follows.
Let $\Phi$ be a stationary point process on $\mathbb R ^2$ (i.e.\ its distribution is invariant under planar translations) with intensity $\kappa>0$, and associate to each point $(\xi,\eta)\in\Phi$ a point process $X_{(\xi,\eta)}\subset\mathbb{R}^3$ that is concentrated around the line in $\mathbb R^3$ which is perpendicular to the $xy$-plane, with intersection point $(\xi,\eta,0)$. We refer to $X_{(\xi,\eta)}$ as the cylindrical cluster associated to $(\xi,\eta)$. Let $P_{xy}(X_{(\xi,\eta)}\cap W)$ denote the projection onto the $xy$-plane of the observed part of the cylindrical cluster. 
For short we refer to the non-empty $P_{xy}(X_{(\xi,\eta)}\cap W)$ as the projected cluster with centre point $(\xi,\eta)$.
Then we let \[X_{xy}=\bigcup_{(\xi,\eta)\in\Phi} P_{xy}(X_{(\xi,\eta)}\cap W).\] 
Further, conditioned on $\Phi$, we assume that the projected clusters are independent and each non-empty $P_{xy}(X_{(\xi,\eta)}\cap W)$ is distributed as the intersection of $W_{xy}$ with a finite planar Poisson process  translated by the centre point $(\xi,\eta)$; this Poisson process has intensity function $a\alpha f$, where $a$ is the length of the interval $W_z$, $\alpha>0$ is a parameter, and $f$ is the probability density function of a bivariate zero-mean isotropic normal distribution with standard deviation $\sigma>0$. Thus, ignoring boundary effects, $\alpha a$ is the expected size (or number of points) of a projected cluster and $\sigma$ controls the spread of points in a projected cluster.    
Specifically, we let first $\Phi$ be a planar stationary Poisson process and later a stationary determinantal point process \citep{Lavancier:etal:2015}, since later we observe for a fitted Poisson process a very low expected number of points in a projected cluster and that there is a need for a repulsive model in order to obtain less overlap between the projected clusters.  %with a standard deviation $\sigma$; thus, in accordance with the minicolumn hypothesis, we may expect  

\subsubsection{The model for $X_z$ conditioned on $X_{xy}$}\label{s:model2}
Consider the special case where $\Phi$ is a planar stationary Poisson process and $X_z$ conditioned on $X_{xy}$ is a homogeneous binomial point process, that is, the $n$ points in $X_z$ are independent and uniformly distributed on $W_z$. Note that $X_z$ depends only on $X_{xy}$ through the number of points in $X_{xy}$.
It becomes clear in Section~\ref{s:dPLCPP} that this special case corresponds to a degenerate case of a Poisson line cluster point process  (PLCPP) as considered in \cite{Moller:etal:2016}.  

In Section~\ref{s:gen_dPLCPP} we  consider several other cases than a homogeneous binomial point process for $X_z$ conditioned on $X_{xy}$. In the end, in comparison with the PLCPP model suggested in \cite{Moller:etal:2016} and \cite{Rafati:etal:2016} for the description of another dataset of pyramidal nucleous locations, we suggest and fit a rather complex hierarchical point process model. Our final model describes columnar structures of the nucleolus locations in each dataset \texttt{L3} and \texttt{L5}, with repulsion between nucleolus locations given by a hard core condition on a small scale and a stunted cylindrical interaction region on a larger scale, as well as clustering between nucleolus locations given by an elongated cylindrical interaction region. In particular, our final fitted model describes a columnar structure but with much smaller columns than expected under the minicolumn hypothesis.

\subsection{Outline}
The remainder of this paper is organised as follows. 
In Section~\ref{s:prelim} we introduce some basic concepts and definitions needed when introducing and fitting the models in the subsequent sections. 
In Section~\ref{s:CSR} we investigate how the nucleolus locations deviate from complete spatial randomness.
In Section~\ref{s:dPLCPP} we also notice a deviation from a fitted degenerate PLCPP model. 
In Section~\ref{s:gen_dPLCPP} we introduce and fit various generalisations of the degenerate PLCPP model, including the final model briefly described in Sections~\ref{s:model1}--\ref{s:model2}. In particular, for estimation of parameters used in models for $X_z$ conditioned on $X_{xy}$ we propose in Section~\ref{s:gen_dPLCPP} a maximum pseudo likelihood procedure, the performance of which is studied in a simulation study reported in Appendix I.
Finally, Section~\ref{s:discussion} summaries our findings and discuss directions for future research.

\section{Preliminaries}\label{s:prelim}
The point processes $X$, $X_{xy}$, and $X_z$ introduced above can be viewed as the restriction to the bounded sets $W$, $W_{xy}$, and $W_z$ of a locally finite point process on $\mathbb R^d$ with $d=3,2,1$,  respectively. Below we recall a few basic statistical tools needed in this paper, using the generic notation $Y$ for a locally finite point process defined on $\mathbb R^d$ (apart from the cases above, we have in mind that $Y$ could also be the centre process $\Phi$ from Section~\ref{s:models}). Briefly speaking, this means that $Y$ is a random subset of $\mathbb{R}^d$ satisfying that $Y_B = Y\cap B$ is finite for any bounded set $B\subset\mathbb{R}^d$; for a more rigorous definition of point processes, see e.g.\ \cite{DVJ:2003} or \cite{MW:2004}.

\subsection{Moments}
For each integer $k\ge1$, to describe the $k$'th order moment properties of $Y$, we consider the so-called $k$'th order intensity function $\lambda^{(k)}:(\mathbb R^d)^k\rightarrow[0,\infty)$ given that it exists. This means that for any pairwise distinct and bounded Borel sets $B_1,\ldots,B_k\subset\mathbb{R}^d$,  
\begin{equation*}
	\text{E} \left[n(Y_{B_1})\cdots n(Y_{B_k})\right]=
	\int_{B_1}\cdots\int_{B_k}\lambda^{(k)}(x_1,\ldots,x_k)\,\mathrm{d}x_1\cdots\,\mathrm{d}x_k
\end{equation*}
is finite, where $n(Y_B)$ denotes the cardinality of $Y_B$.

The first order intensity function $\lambda^{(1)}=\lambda$ is of particular interest and is simply referred to as the intensity function. Heuristically, $\lambda(u)\,\mathrm{d}u$ can be interpreted as the probability of observing a point from $Y$ in the infinitesimal ball of volume $\mathrm{d}u$ centred at $u$. If the intensity function $\lambda(\cdot)\equiv\lambda$ is constant, then $\lambda|B| = \text{E}\left[n(Y_B)\right]$ for any bounded Borel set $B\subset \mathbb{R}^d$, where $|\cdot|$ is the Lebesgue measure. In this case $Y$ is said to be homogeneous and otherwise inhomogeneous.
Clearly, stationarity of $Y$ (meaning that its distribution is invariant under translations in $\mathbb R^d$) implies homogeneity.

\subsection{Functional summaries}\label{sec:def}
The functional summaries described in this section %Section~\ref{sec:def}
will be used both for model fitting as described in Section~\ref{s:fitting_procedures} and for model checking as described in Section~\ref{sec:GERL}.

To summarise the second order moment properties, it is custom to consider the pair correlation function (PCF), $g$, which is defined as the ratio of the second and first order intensity function, that is,
\begin{equation*}
	g(x_1, x_2) = \frac{\lambda^{(2)}(x_1, x_2)}{\lambda(x_1)\lambda(x_2)}, \quad x_1, x_2\in\mathbb{R}^d.
\end{equation*}
Heuristically, $\lambda(x_1)\lambda(x_2)g(x_1,x_2)$ is the probability of simultaneously observing a point from $X$ in each of the two infinitesimal balls of volume $\mathrm{d}x_1$ and $\mathrm{d}x_2$ centred at respectively $x_1$ and $x_2$.
For a Poisson process $Y$, $g=1$.
The PCF is said to be stationary when (with abuse of notation) $g(x_1, x_2) = g(x_1-x_2)$; this is the case when $Y$ is stationary. 

If the PCF is stationary, it is closely related to the so-called second order reduced moment measure, $\mathcal{K}$, given by
\begin{equation*}
	\mathcal{K}(B) = \int_{B}g(x)\,\mathrm{d}x,
\end{equation*}
where $B\subset\mathbb{R}^d$ is a Borel set \citep[see][]{MW:2004}.
If $Y$ is stationary, % and $B$ has centre of mass at the origin of $\mathbb R^d$, 
then $\lambda\mathcal{K}(B)$ can be interpreted as the expected number of points in $Y\setminus\{o\}$ within $B$ given that $Y$ has a point at the origin $o$ of $\mathbb R^d$; when considering scalings of $B$, we refer to $B$ as a structuring element.
The simplest example occurs when $B$ is a ball centred at the origin and with radius $r>0$; then $K(r)=\mathcal K(B)$ becomes the $K$-function introduced by \cite{Ripley:1976}; and 
often we instead consider a transformation of the $K$-function, which is called the $L$-function and defined by $L(r) = (K(r)/\omega_d)^{1/d}$, where $\omega_d$ is the volume of the $d$-dimensional unit ball. In particular, if $Y$ is a stationary Poisson process, then $L(r)=r$.

For detecting cylindrical structures, \cite{Moller:etal:2016} introduced the cylindrical $K$-function which corresponds to $\mathcal{K}(B)$ when $B$ is a cylinder of height $2t$, base-radius $r$, and centre of mass at the origin. 
Note that Ripley's $K$-function depends only on one argument, $r$, while the cylindrical $K$-function depends both on $r$, $t$, and the direction of the cylinder.
However, when $d=3$ and since the minicolumns are expected to extend along the $z$-axis, we only consider cylinders extending in this direction, effectively reducing the number of arguments to two.

We will also consider the commonly used $F$-, $G$-, and $J$-function when performing model control; see \cite{Lieshout:Baddeley:1996} for definitions. Briefly, if $Y$ is stationary, $F(r)$ is the probability that $Y$ has a point within distance $r>0$ from an arbitrary fixed location in $\mathbb R^d$; $G(r)$ is the probability that $Y$ has another point within distance $r>0$ from an arbitrary fixed point in $Y$; and $J(r)=(1-G(r))/(1-F(r))$ when $F(r)<1$.  

\subsection{Model fitting} 
\label{s:fitting_procedures}
In \citet{Moller:etal:2016} parameter estimation for the degenerate PLCPP model was simply done by a moment based procedure which included minimisation of a certain contrast between a theoretical second order moment functional summary and its empirical estimate. 
Below we describe a similar minimum contrast procedure for estimating the parameters of models for $X_{xy}$. For the models of $X_z$ conditioned on $X_{xy}$ we find it convenient to use a maximum pseudo likelihood procedure as detailed in Section~\ref{sec:MPLE}.

Minimum contrast estimation is a computationally simple fitting procedure introduced by \cite{Diggle:et:al:1984} that is applicable when a closed form expression of a functional summary, $T$, exists. 
The idea is to minimise the distance from the theoretical function $T$ to its empirical estimate $\hat{T}$ for the data. Specifically, if $T$ depends on the parameter vector $\theta$ and is a function of `distance' $r>0$ (as for example in case of Ripley's $K$-function), the minimum contrast estimate of $\theta$ is given by
\begin{equation}\label{e:mc}
\hat{\theta} = \text{argmin}_{\theta}\int_{r_{\text{min}}}^{r_{\text{max}}}\left|T(\theta, r)^q - \hat{T}(r)^q\right|^p\mathrm{d}r,
\end{equation}
where $0\le r_{\text{min}} < r_{\text{max}}$, $q>0$, and $p>0$ are tuning parameters.
General recommendations on $q$ are given in \cite{Guan:2009} and \cite{D:2014}, when $T(r) = g(r)$ or $T(r) = K(r)$; when fitting a model to $X_{xy}$, we let $p=2$, $q=1/4$, $r_{\text{min}} = 0$, and $r_{\text{max}}$ be one fourth of the shortest side length of $W_{xy}$.

When the PCF has a closed form expression, alternative estimation procedures can be used, including the second order composite likelihood \citep[see][]{Guan:2006,Waagepetersen:2007}, adapted second order composite likelihood \citep[see][]{Lavancier:etal:2018}, and Palm likelihood \citep[see][]{Ogata:etal:1991,Prokesova:etal:2016,BRT:2016}.

\subsection{GERL envelope procedure}\label{sec:GERL}
For model checking we consider informative global extreme rank length (GERL) envelope procedures \citep{Mrkvicka:etal:2018,Myllymaki:etal:2017} based on various functional summaries  as described below. 

In the GERL envelope procedure, the distribution of an empirical functional summary is estimated by simulations under a fitted model of interest.
The procedure is a refinement of the global rank envelope procedure \citep{Myllymaki:etal:2017}, where it is recommended to use 2499 simulations for a single one-dimensional functional summary and at least 9999 simulations for a single two-dimensional functional summary \citep{Mrkvicka:etal:2016}. However, we consider a concatenation of the $L$-, $G$-, $F$-, and $J$-function in which case \cite{Mrkvicka:etal:2017} recommend using more simulations. Particularly for a concatenation of $k$ one-dimensional summary functions they recommend using $k\times 2499$ simulations. 
For the GERL envelope procedure, \cite{Mrkvicka:etal:2018} suggest that a lower number of simulations may be enough. Therefore, we use 9999 simulations.
Since we consider a concatenation of one-dimensional functional summaries, we ensure that each of the functional summaries are weighted equally in the GERL envelope test by evaluating them at the same number of arguments \citep{Mrkvicka:etal:2017}. Specifically, we consider 64 $r$-values for each of the functions $L$, $G$, $F$, and $J$. 
The cylindrical $K$-function is not part of the concatenation, and it will be evaluated over a square grid consisting of 64 $r$-values and 64 $t$-values.

\subsection{Summary}
We remark that the summary function used for model fitting will not be used for checking goodness of fit using the GERL envelope procedure. In summary, we use  
\begin{itemize}
	\item empirical estimates of the cylindrical $K$-function for investigating anisotropy and in particular columnarity in the 3D point patterns; 
	\item when considering isotropic point process models for the 2D projected locations (projections into the two-dimensional $xy$-plane), Ripley's $K$-function for parameter estimation (both the theoretical $K$-function and its parametric estimate are used as explained in Section~\ref{s:fitting_procedures}) and empirical estimates of the $G$-, $F$-, and $J$-function 
	when checking for goodness of fit;
	\item when checking for goodness of fit for anisotropic point process models fitted to the 3D point patterns, empirical estimates of both the $L$-, $G$-, $F$-, and $J$-function (which are not informative about anisotropy) and the cylindrical $K$-function (which is informative about anisotropy).   
\end{itemize}

\section{Complete spatial randomness}\label{s:CSR}
The most natural place to begin our point pattern analysis is by testing whether a homogeneous Poisson process $X$ with intensity $\lambda>0$ (we then view $Y$ as a stationary Poisson process with the same intensity), also called complete spatial randomness (CSR), adequately describe each nucleolus
point pattern dataset. % $x\subset W$.
Recall that this means that
%a point process $Y$ is specified such that, for any bounded Borel set $B\subset\mathbb{R}^3$, 
$n(X)$ is Poisson distributed with parameter $\lambda|W|$ and conditional on $n(X)$ the points in $X$ are independent and uniformly distributed within $W$.
We shall see that CSR is a too simple model for the description of \texttt{L3} and \texttt{L5}, and that the noticed deviations from CSR  
become useful for suggesting new models.

\begin{figure}
	\centering
	\includegraphics[width=0.49\linewidth, trim={14mm 52mm 6.5mm 60mm}, clip]{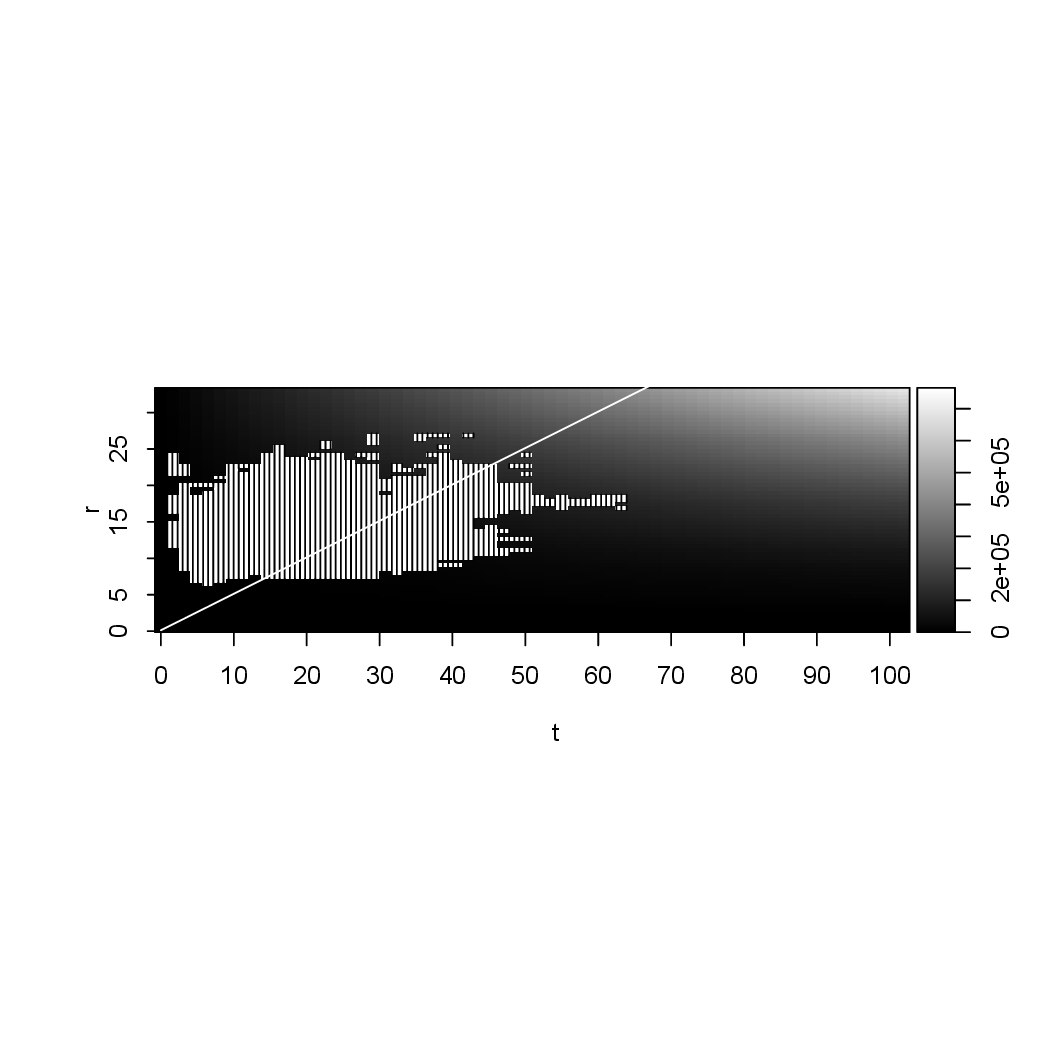}%
	\includegraphics[width=0.49\linewidth, trim={14mm 52mm 6.5mm 60mm}, clip]{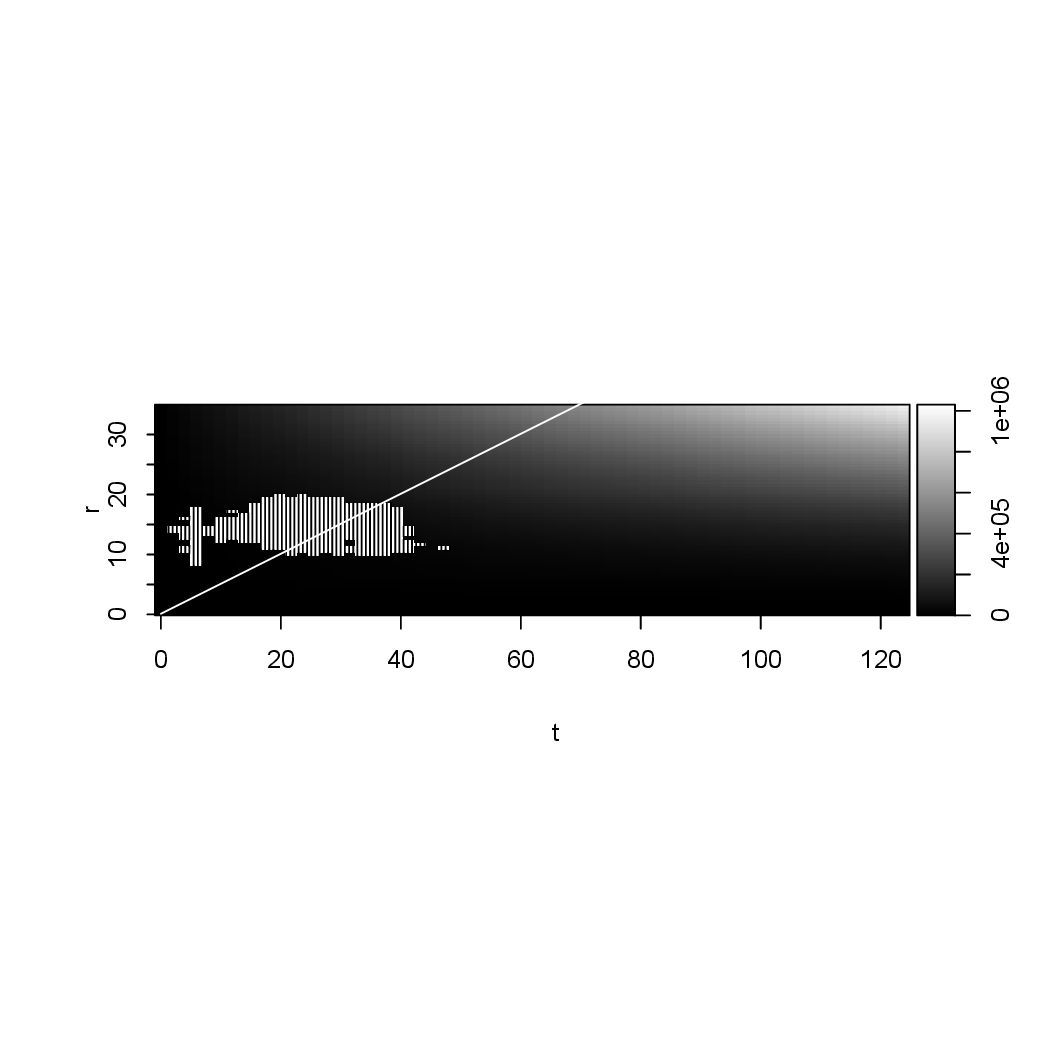}%
	\\ 
	\includegraphics[width=0.49\linewidth, trim={14mm 52mm 6.5mm 60mm}, clip]{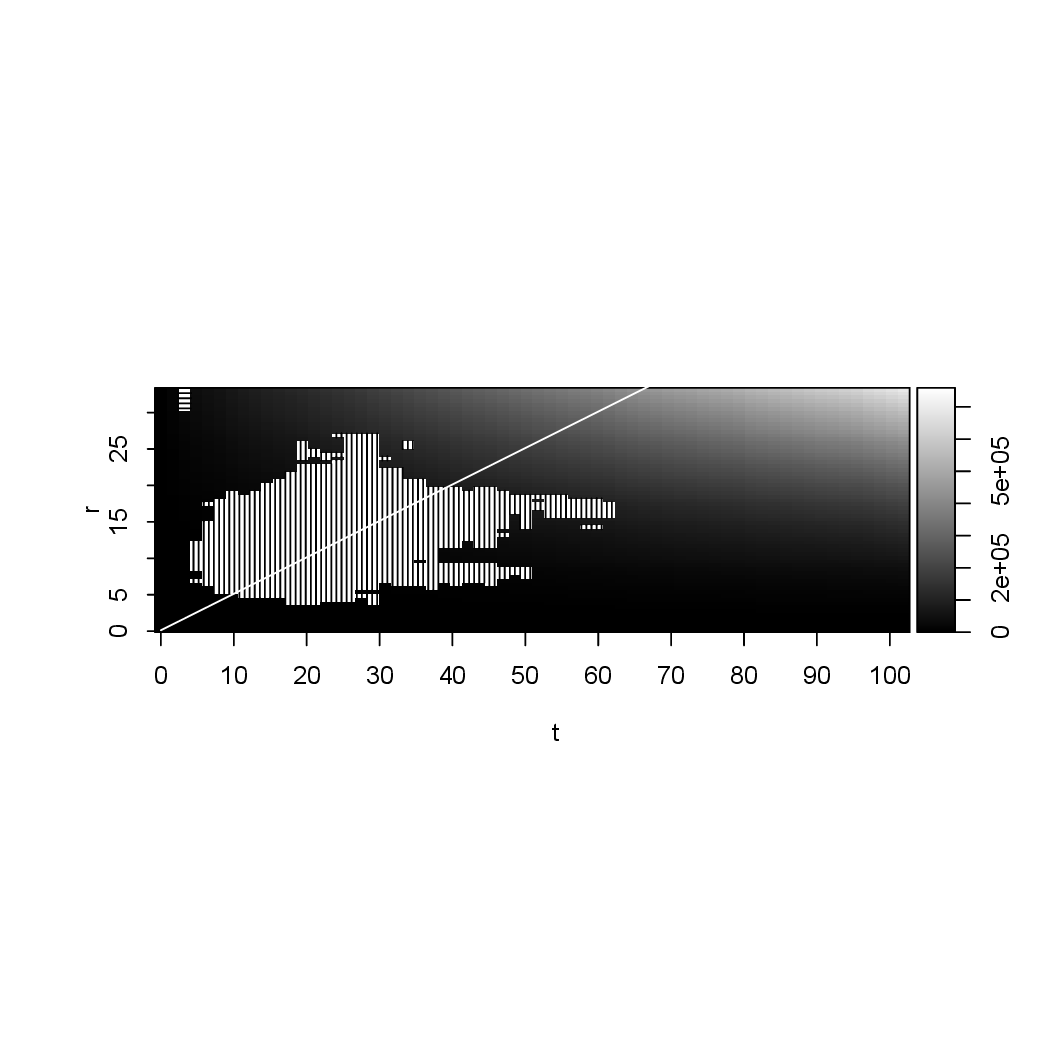}%
	\includegraphics[width=0.49\linewidth, trim={14mm 52mm 6.5mm 60mm}, clip]{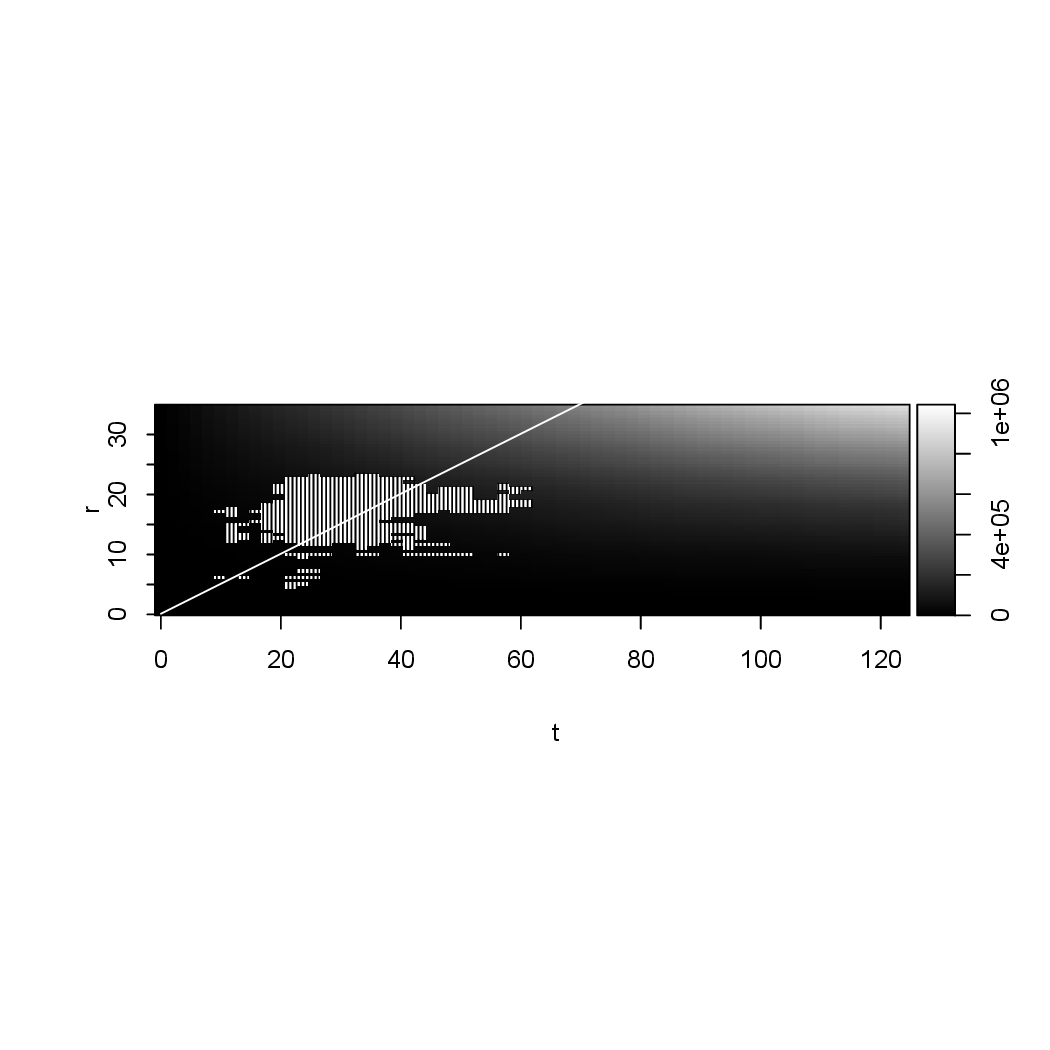}%
	\\
	\includegraphics[width=0.49\linewidth, trim={14mm 52mm 6.5mm 60mm}, clip]{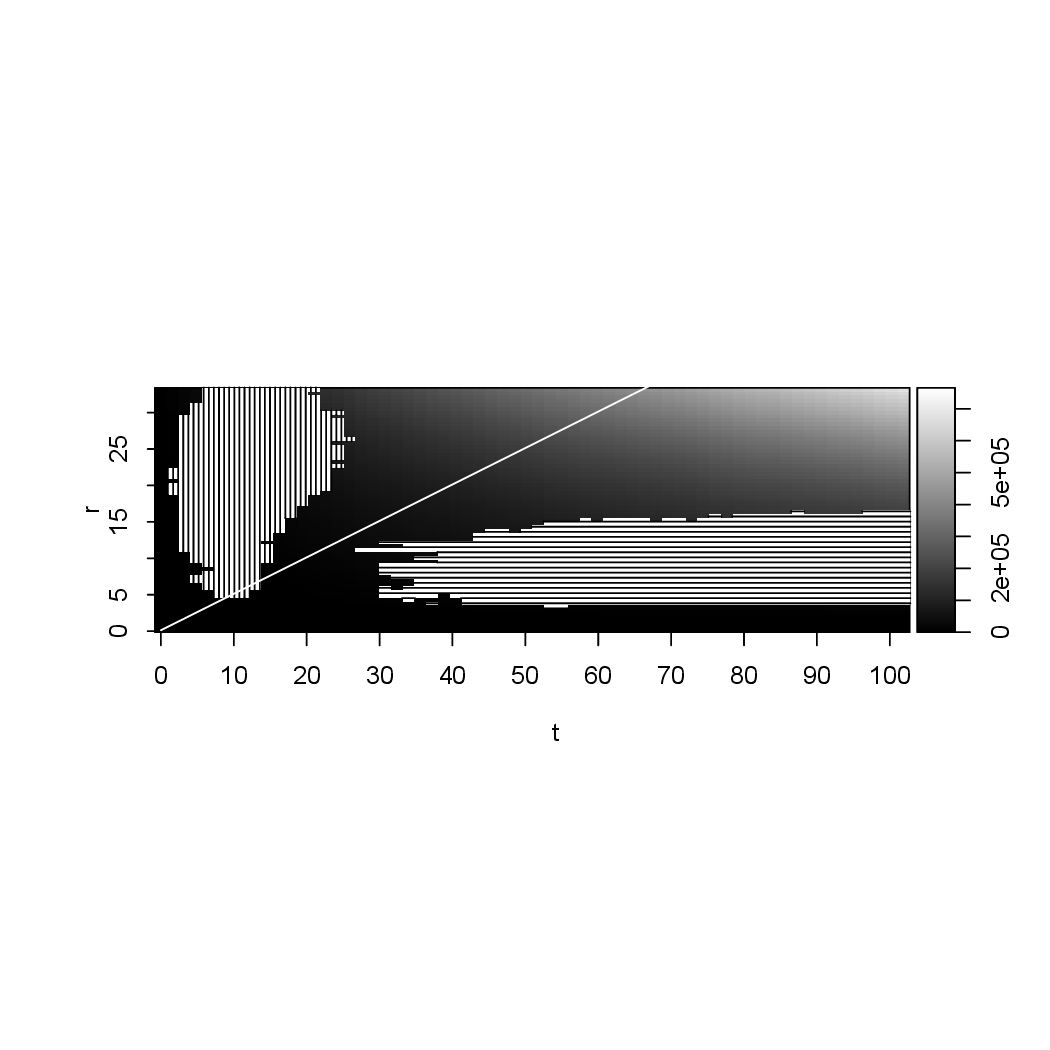}%
	\includegraphics[width=0.49\linewidth, trim={14mm 52mm 6.5mm 60mm}, clip]{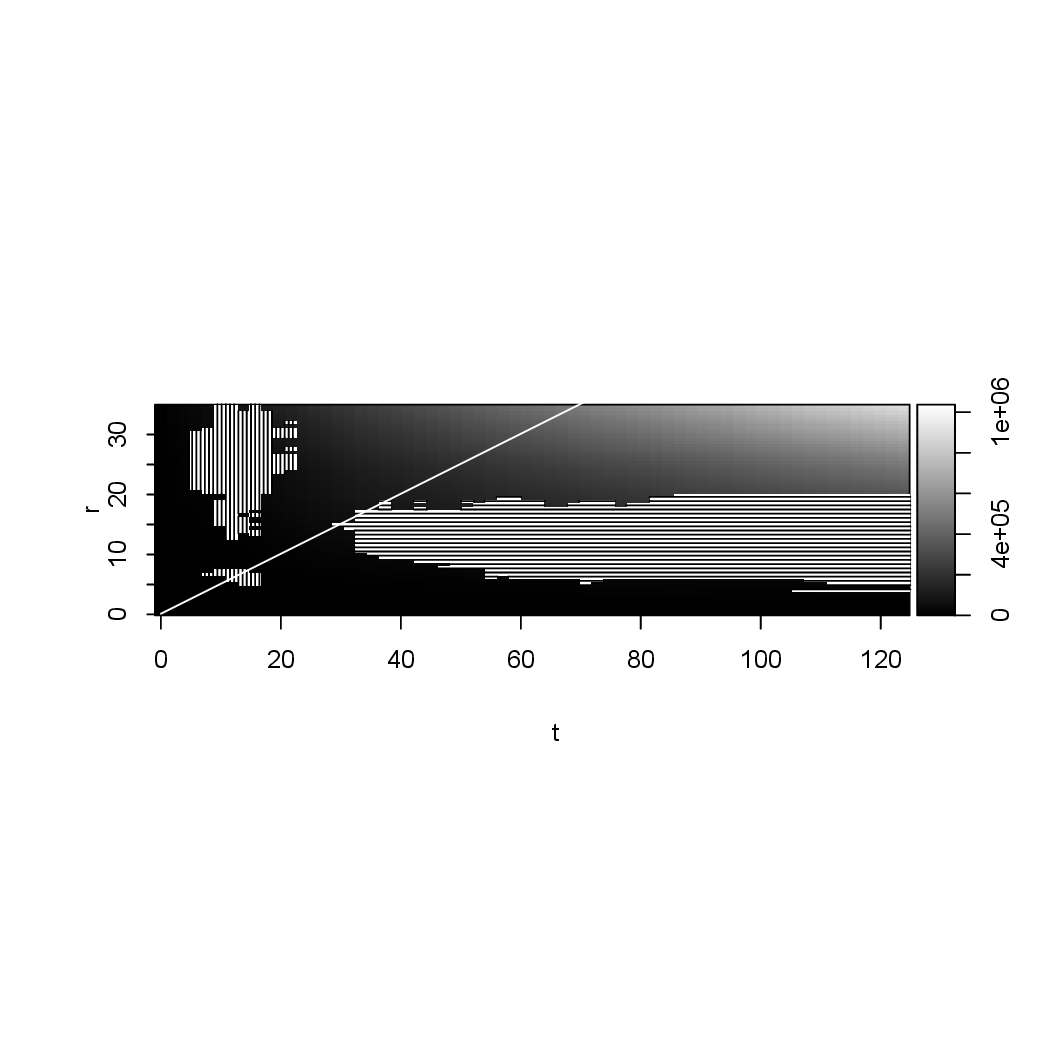}%
	\caption{{Results of the GERL envelope procedure under CSR based on the empirical cylindrical $K$-function (grey scale) with the cylindrical structuring element extending in the $x$-, $y$-, and $z$- directions (top, middle, and bottom, respectively). The shaded horizontal/vertical lines indicate that the function falls above/below the 95\% GERL envelope. The white line indicates the values for which the cylinder height is equal to the base diameter. Left: results for the dataset \texttt{L3}. Right: results for the dataset \texttt{L5}.}}
	\label{f:summ_fun_CSR}
\end{figure}

The CSR model is fully specified by its intensity, which naturally is estimated by $n(X)/|W|$, which is equal to $2.37\times 10^{-5}$ for \texttt{L3} and $1.63\times 10^{-5}$ for \texttt{L5}. 
For that fitted model, Figure~\ref{f:summ_fun_CSR} summarises the results of the GERL envelope procedure, based on the empirical cylindrical $K$-function with the cylindrical structuring element extending in the $x$-, $y$-, and $z$-directions, along with the areas at which the empirical cylindrical $K$-function falls outside the GERL $95\%$ envelope.
It is clearly seen that CSR is rejected for both point patterns in all three directions. This is supported by the associated $p$-values of $10^{-4}$ for all three directions for \texttt{L3}, and $4*10^{-4}$ for the $x$ and $y$-directions and $10^{-4}$ for the $z$-direction for \texttt{L5}.
We notice that the empirical cylindrical $K$-function extending in the $z$-direction falls above the upper GERL envelopes for cylinders that have a height larger than approximately 35\,$\mu\mathrm{m}$ for both datasets, together  with a base radius of approximately 5--15\,$\mu\mathrm{m}$ for \texttt{L3} and 5--20\,$\mu\mathrm{m}$ for \texttt{L5}. This trend is not observed for the empirical cylindrical $K$-functions extending in the $x$- and $y$-directions.
Furthermore, the observed cylindrical $K$-functions extending in the $z$-direction falls below the lower $95\%$ GERL envelope for cylinders with a height of approximately 10--25\,$\mu\mathrm{m}$ and a base radius larger than 5\,$\mu\mathrm{m}$.
We further observe that the empirical cylindrical $K$-functions extending in the $x$- and $y$- directions falls below the lower $95\%$ GERL envelope for cylinders with a height of approximately 0--60\,$\mu\mathrm{m}$ and a base radius of approximately 5--25\,$\mu\mathrm{m}$.
Hence, for elongated cylinders extending in the $z$-direction, we tend to see more points in the data than we expect under CSR, while for stunted cylinders we tend to see fewer points. {Similarly, for cylinders that are neither very stunted or very elongated in the $x$- and $y$-directions, we see fewer points than we expect to see under CSR.}
This seems to be in correspondence with columnar structures where the columns extend in the $z$-direction.

\section{The degenerate Poisson line cluster point process}\label{s:dPLCPP}
\cite{Moller:etal:2016} presented the so-called Poisson line cluster point process (PLCPP) which is useful for modelling columnar structures. Specifically, we consider a degenerate PLCPP $Y \subset \mathbb{R}^3$ constructed as follows.
\begin{enumerate}
	\item\label{i:def_PLCPP_PLP} Generate a stationary Poisson process $\Phi=\{(\xi_i, \eta_i) \}_{i = 1}^{\infty}\subset \mathbb{R}^2$ with finite intensity $\kappa>0$. Each point $(\xi_i, \eta_i)\in \Phi$ corresponds to an infinite line $l_i$ in $\mathbb{R}^3$ which is perpendicular to the $xy$-plane, that is, $l_i = \left\{(\xi_i, \eta_i, z)\, |\, z\in\mathbb{R} \right\}$. 
	\item\label{i:def_PLCPP_PP} Conditional on $\Phi$, generate independent stationary Poisson processes $L_1\subset l_1, L_2\subset l_2, \ldots$ with identical and finite intensity $\alpha>0$.
	\item\label{i:def_PLCPP_clusters} Generate point processes  $X_1, X_2, \ldots \subset \mathbb{R}^3$ by independently displacing the points of $L_1, L_2, \ldots$ by the zero-mean isotropic normal distribution with standard deviation $\sigma>0$.
	\item\label{i:def_PLCPP_super} Finally, set $Y = \bigcup_{i=1}^{\infty} X_i$ and $X=Y_W$.
\end{enumerate}

Some comments to the construction in items~\ref{i:def_PLCPP_PLP}--\ref{i:def_PLCPP_super} are in order.

In the general definition of the PLCPP in \cite{Moller:etal:2016}, the lines $l_1, l_2, \ldots$ are modelled as a stationary Poisson line process. That is, the lines are not required to be perpendicular to the $xy$-plane nor does the Poisson line process need to be degenerate (meaning that the lines are not required to be mutually parallel). Further, the dispersion density (used in item~\ref{i:def_PLCPP_clusters}) can be arbitrary. However, the construction is still such that $Y$ becomes stationary. Furthermore, 
the same distribution of $Y$ is obtained whether we consider a three-dimensional zero-mean isotropic normal distribution for displacements in item~\ref{i:def_PLCPP_clusters} or a bivariate zero-mean isotropic normal distribution with displacements of the $xy$-coordinates for the points of $L_1, L_2, \ldots$, provided the variances of the two normal distributions are identical, cf.\ \cite{Moller:etal:2016}. 

Returning to the degenerate PLCPP of items~\ref{i:def_PLCPP_PLP}--\ref{i:def_PLCPP_super}, we imagine  that each $X_i$ is a cylindrical cluster of points around the line $l_i$, where these cylindrical clusters are  parallel to the $z$-axis. Furthermore, the interpretation of the parameters $\kappa$, $\alpha$, and $\sigma$ in terms of a Poisson cluster point process is similar to that in Section~\ref{s:model1} except that we now also consider lines not intersecting $W$: if $Y$ as defined by items~\ref{i:def_PLCPP_PLP}--\ref{i:def_PLCPP_super} is restricted to a subset $S\subset\mathbb{R}^3$ bounded by two planes parallel to the $xy$-plane, for specificity $S=\left\{(x, y, z)\in \mathbb{R}^3\,|\,z\in W_z\right\}$, this restricted point process can be seen as a (modified) Thomas process \citep[see][]{Thomas:1949, MW:2004} on $\mathbb{R}^2$ along with independent $z$-coordinates following a uniform distribution on $W_z$. 

To see this, first note that conditional on $\Phi = \{(\xi_i, \eta_i) \}_{i = 1}^{\infty}$ and for all $i = 1, 2, \ldots$, $X_i$ is a Poisson process in $\mathbb{R}^3$ with intensity function $\lambda_i((x, y, z)) = \alpha f(x-\xi_i, y-\eta_i)$, where $f$ is the probability density function of the bivariate isotropic normal distribution given in item~\ref{i:def_PLCPP_clusters}. In turn, this implies that $Y$ conditioned on $\Phi$ is a Poisson process in $\mathbb{R}^3$ with intensity function $\sum_{i = 1}^{\infty}\lambda_i((x, y, z))$. Further, since $\lambda_i(x, y, z) = \lambda_i(x, y)$ does not depend on $z$ for all $i=1, 2, \ldots$, the projection of $Y_S$ onto the $xy$-plane, $P_{xy}(Y_S)$, conditioned on $\Phi$ is a Poisson process with intensity $a\sum_{i = 1}^{\infty}\lambda_i(x, y)$, where $a$ is the length of the interval $W_z$. Since $\Phi$ is a stationary Poisson process, $P_{xy}(Y_S)$ is a Thomas process with centre process intensity $\kappa$ and expected cluster size $\alpha a$ (that is, the expected number of points in $X_i\cap S$). Finally, from items~\ref{i:def_PLCPP_PP}--\ref{i:def_PLCPP_super} it follows that the $z$-coordinates of $X_z$ are independent and uniformly distributed on $W_z$, and they are independent of $X_{xy}$.

Consequently, simulating $X=Y_W$ is straightforwardly done by simulating a Thomas point process (on a larger set than $W_{xy}$ in order to avoid boundary effects) along with independent uniform $z$-coordinates on $W_z$. For simulating the Thomas point process we apply standard software from the \texttt{R}-package \texttt{spatstat} \citep{BRT:2016}. Similarly,
fitting a degenerate PLCPP to a realisation of $X$ is simply a matter of fitting a Thomas process to the point pattern consisting of the $xy$-coordinates of the points in that realisation. Since the $K$-function of the Thomas process has a closed form expression, the model can easily be fitted using minimum contrast estimation with $T(r) = K(r)$ in \eqref{e:mc}.
Table~\ref{t:MC_PLCPP} summarises the parameter estimates, where most notably the expected cluster size $\widehat{\alpha a}$ is $< 1$ for both \texttt{L3} and \texttt{L5}. Understanding each cylindrical cluster within $W$ as (a part of) a minicolumn, `these parameter estimates result in very unnatural models for the datasets, since each minicolumn within $W$ is expected to consist of less than one point' (personal communication with Jens R. Nyengaard).
\begin{table}[!ht]
	\centering
	\begin{tabular}{l|lll}
		& $\hat{\kappa}$ & $\hat{\sigma}$ & $\widehat{\alpha a}$ \\ \hline
		\texttt{L3} & 0.027 & 2.86 & 0.36 \\
		\texttt{L5} & 0.0085 & 4.58 & 0.95 \\
	\end{tabular}
	\caption{Minimum contrast estimates of the degenerate PLCPP.}
	\label{t:MC_PLCPP}
\end{table}

Despite the fact that the fitted degenerate PLCPP models are somewhat unnatural and hardly can be interpreted as a model with (the hypothesised) minicolumnar structures, GERL envelope procedure based on a concatenation of the $F$-, $G$-, and $J$-function shows that the Thomas process suitably fit the projected locations {(projections into the two-dimensional $xy$-plane)} with a $p$-value of 0.76 for \texttt{L3} and 0.87 for \texttt{L5}. 
However, results from the concatenated GERL envelope procedure described in Section~\ref{sec:GERL} indicated that the model did not suitably describe the three-dimensional nucleolus locations with $p$-values of $10^{-4}$ and $10^{-4}$ for \texttt{L3} and $10^{-4}$ and $6*10^{-4}$ for \texttt{L5} when using the GERL envelope procedure based on the concatenation of the one-dimensional summary functions ($L-r$, $G$, $F$, $J$) and the cylindrical $K$-function in the $z$-direction, respectively.
\begin{figure}
	\centering
	\includegraphics[width=0.7\linewidth, trim={14mm 52mm 6.5mm 60mm}, clip]{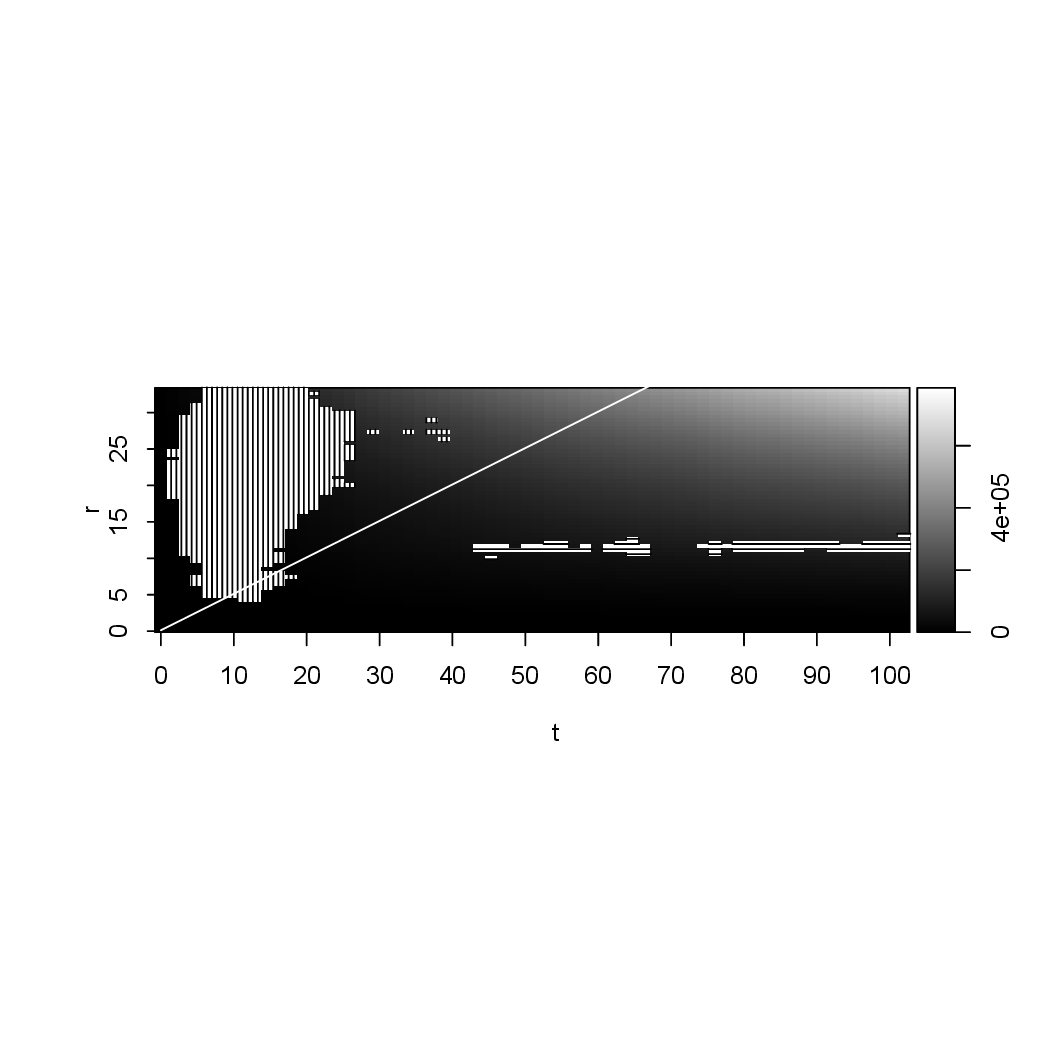}
	\\[0.5em]
	\includegraphics[width=0.7\linewidth, trim={14mm 52mm 6.5mm 60mm}, clip]{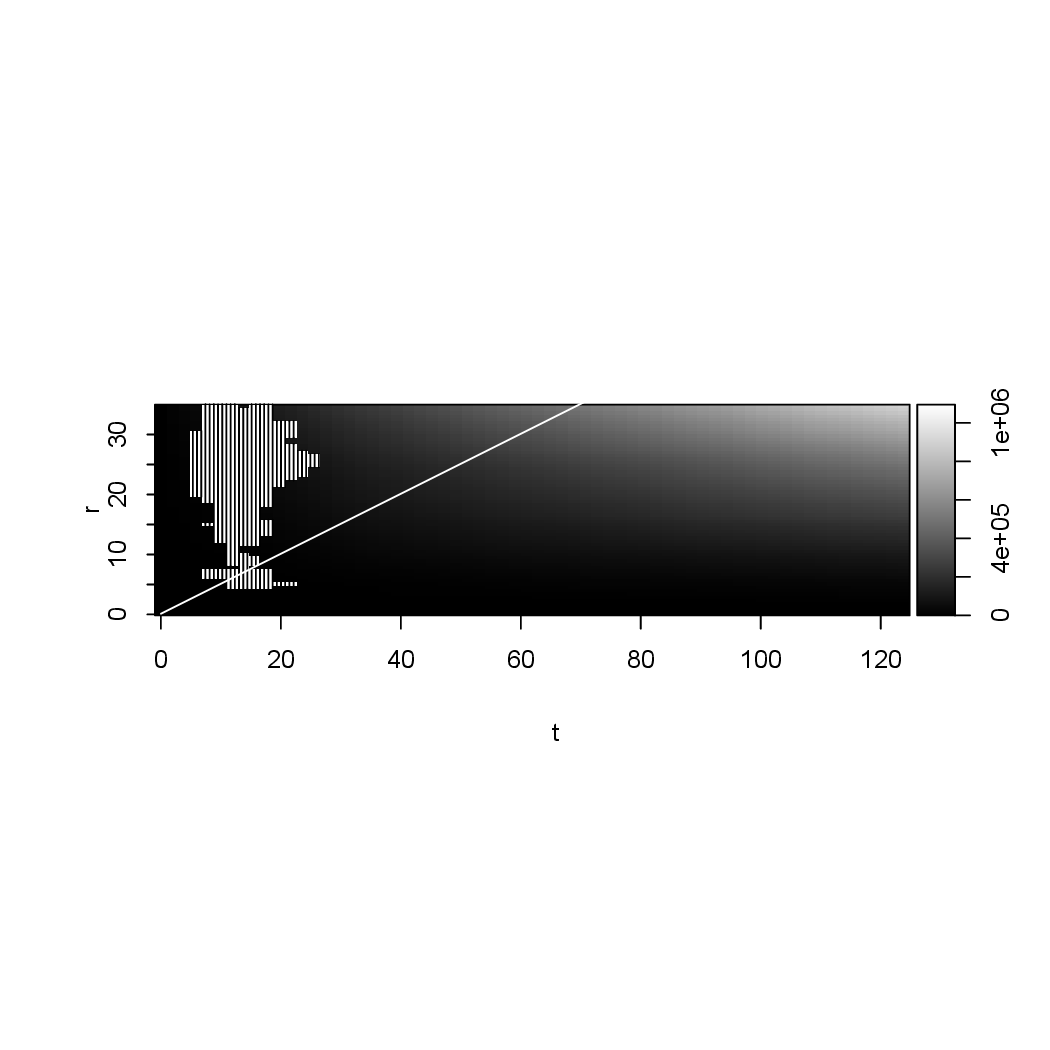}
	\caption{Empirical estimates of the cylindrical $K$-function (grey scale) where shaded horizontal/vertical lines indicate that the function falls above/below the 95\% GERL envelope under the fitted degenerate PLCPP.
	The white line indicates the values for which the cylinder height is equal to the base diameter. Top: results for the dataset \texttt{L3}. Bottom: results for the dataset \texttt{L5}.}
	\label{f:summ_fun_PLCPP}
\end{figure}
Specifically, Figure~\ref{f:summ_fun_PLCPP} shows the empirical cylindrical $K$-function and indicates where it deviates from the 95\% GERL envelope. Clearly, the model does account for some of the hypothesized columnarity of the data as opposed to CSR, but the empirical cylindrical $K$-function for \texttt{L3} still falls above the 95\% GERL envelope. Furthermore, the empirical cylindrical $K$-function for both datasets falls below the 95\% GERL envelope similar to what was seen under CSR, indicating a lack of regularity {of the model}, which in fact is supported by the one-dimensional functional summaries (not shown). This could suggest that the cylindrical clusters should be more distinct; motivating us to generalise the degenerate PLCPP model as in the following section.

\section{A generalisation of the degenerate PLCPP}\label{s:gen_dPLCPP}
As some but not all features of the data were explained by the degenerate PLCPP fitted in Section~\ref{s:dPLCPP}, we propose in this section two generalisations as follows.
\begin{enumerate}
	\item The centre process $\Phi$ is a planar stationary point process.  \label{i:generalise_line_process}
	\item $X_z$ conditioned on $X_{xy}$ is a Markov random field.
	\label{i:generalise_to_repulsiveness}
\end{enumerate}

The first modification is straightforward and {although the Thomas point process suitably fitted the projected point patterns we found the parameter estimates unnatural for describing cylindrical clustering as discussed in Section~\ref{s:dPLCPP}. Therefore we chose a repulsive centre process in order to obtain more distinguishable cylindrical clusters;} this is detailed in Section~\ref{s:generalise_line_process}. Further, the assumption of stationarity of $\Phi$ is made in order to apply a similar minimum contrast estimation procedure as in Section~\ref{s:dPLCPP}, so implicitly we make the assumption that the PCF or the $K$-function is expressible in a closed form. 
For the second modification we suggest a conditional model inspired by the multiscale point process and particularly the Strauss hard core point process \citep[see e.g.][]{MW:2004} which will allow for further repulsion or even aggregation between the points; this is detailed in Sections~\ref{s:generalise_to_repulsiveness}--\ref{sec:5.3}.

\subsection{A determinantal point process model for the centre points}\label{s:generalise_line_process}
Consider a point process $Y\subset\mathbb R^3$ specified by items~\ref{i:def_PLCPP_PLP}--\ref{i:def_PLCPP_super} in Section~\ref{s:dPLCPP} with the exception that the centre process $\Phi$ now is an arbitrary stationary planar point process. Then, recalling the notation from Section~\ref{s:dPLCPP}, $P_{xy}(Y_S)$ is a planar Cox process \citep[see][]{MW:2004} and even a planar generalised shot-noise Cox process \citep[see][]{Moller:Torrisi:2005} driven by the random intensity function $\Lambda(x,y)=a\sum_{i=1}^{\infty}\lambda_i(x, y)$ for $(x,y)\in\mathbb R^2$.
Moreover, $P_{xy}(Y_S)$ corresponds to the Thomas process, but with a different centre point process (unless of course $\Phi$ is a stationary Poisson process).

In this section we focus on the case where $\Phi$ is a stationary determinantal point process \citep[DPP; see][]{Lavancier:etal:2015}, in which case we will refer to $Y$  as the determinantal line cluster point process (DLCPP). 
Let $C:\mathbb R^2\times\mathbb R^2\rightarrow \mathbb{C}$ be a function, then $\Phi$ is a DPP with kernel $C$ if its intensity functions satisfy
\begin{equation*}
	\lambda^{(k)}(u_1,\ldots,u_k) = \text{det}[C](u_1, \ldots, u_k)\qquad \mbox{for $k=1, 2, \ldots$, $u_1,\ldots,u_k\in\mathbb R^2$},
\end{equation*}
where $\text{det}[C](u_1, \ldots, u_k)$ is the determinant of the $k\times k$ matrix with $(i,j)$'th entry $C(u_i, u_j)$.
For further details on DPPs, we refer to \cite{Lavancier:etal:2015} and the references therein.
When $\Phi$ is a DPP, we call $P_{xy}(Y_S)$ a determinantal Thomas point process (DTPP). 
The DTPP is discussed to some extent in \cite{Moller:Christoffersen:2018}, where a closed form expression of its PCF is found. 
Thus, the DLCPP can be fitted by fitting a DTPP to the projected data using a minimum contrast procedure (see Section~\ref{s:fitting_procedures}).

Specifically, we let $\Phi$ be  the jinc-like DPP given by the kernel $C(u_1, u_2) = \sqrt{\kappa/\pi}J_1\left(2\sqrt{\pi\kappa}\|u_1-u_2\|\right) / \|u_1-u_2\|$, where $\kappa>0$ is the intensity of $\Phi$, $J_1$ is the first order Bessel function of the first kind, and $\|\cdot\|$ denotes the usual planar distance. In the sense of \cite{Lavancier:etal:2015}, this is the most repulsive DPP with a stationary kernel \citep[see also][]{Biscio:etal:2016}.
Simulation of the DTPP is done by first simulating $\Phi$ on a larger region than $W_{xy}$ in order to avoid boundary effects, for which we use the functionality of \texttt{spatstat}; second we generate for each cluster a Poisson distributed number of points with intensity $\alpha a$; and third we displace these points by a bivariate zero-mean isotropic normal distribution.

\begin{figure}
	\centering
	\includegraphics[scale=0.35, trim={2mm 60mm 2mm 62mm}, clip]{L3_XY.eps}
	\includegraphics[scale=0.35, trim={2mm 60mm 2mm 62mm}, clip]{L5_XY.eps} \\
	\includegraphics[scale=0.35, trim={2mm 58mm 0mm 62mm}, clip]{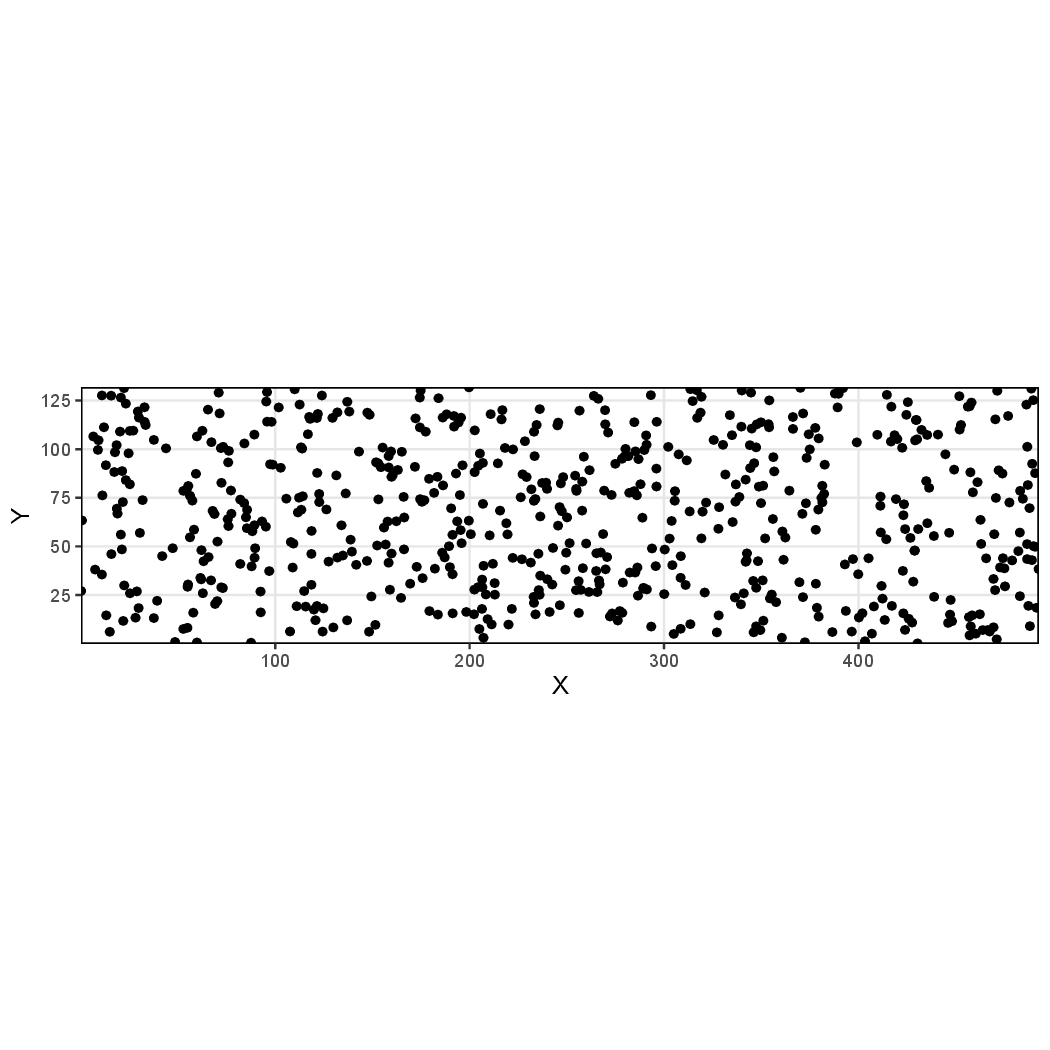}
	\includegraphics[scale=0.35, trim={2mm 58mm 0mm 62mm}, clip]{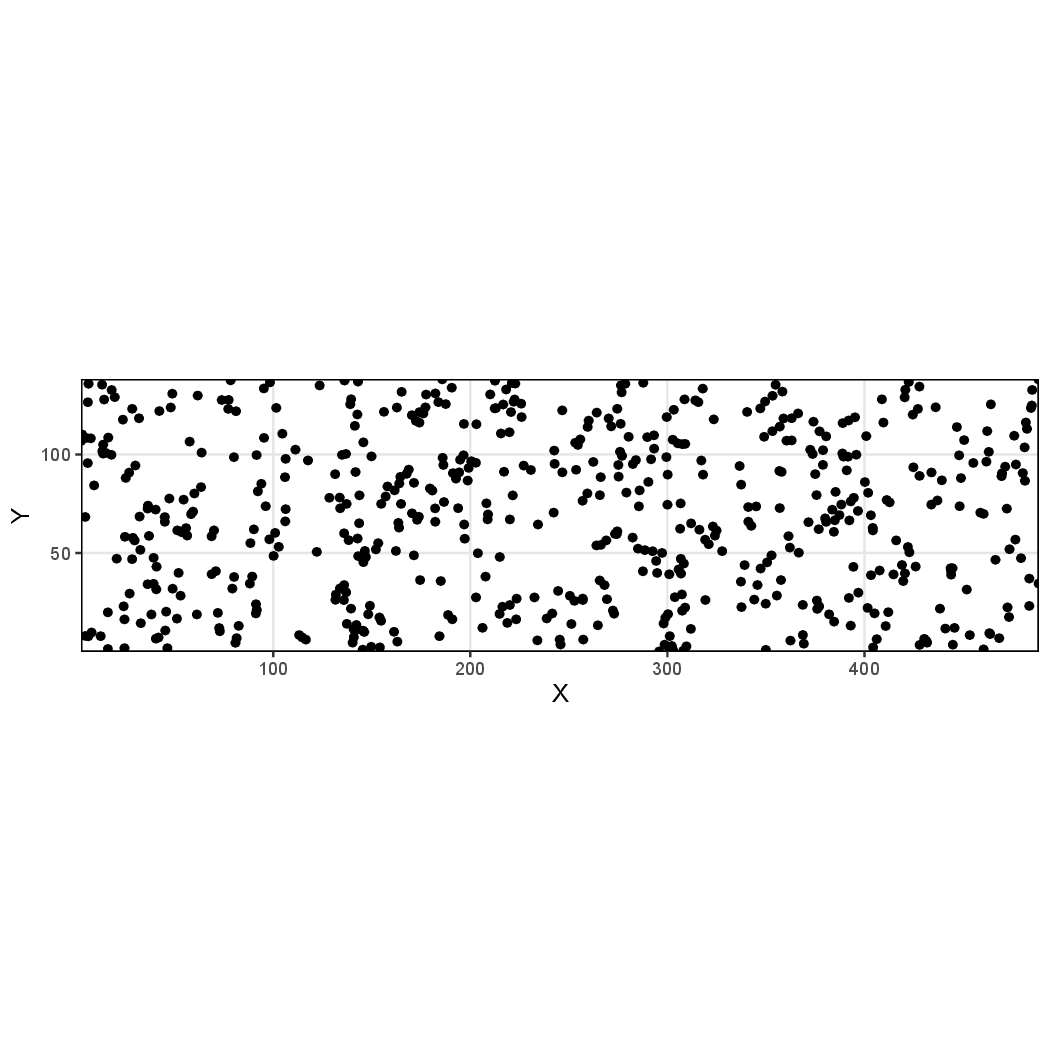}
	\caption{
		Projection of observed nucleolus locations onto the $xy$-plane (left) and simulations from the fitted jinc-like DTPP (right) for the datasets \texttt{L3} (top) and \texttt{L5} (bottom).
	}
	\label{f:pp_proj_and_sim_jDPP}
\end{figure}

The parameter estimates of the jinc-like DTPP model were obtained by minimum contrast with $T(r)=g(r)$. 
The parameter estimates are given in Table~\ref{t:MC_DLCPP}, where the estimated expected cluster size $\widehat{\alpha a}$ is `much smaller than expected for a minicolumn when restricting it to the observation window -- provided the minicolumn hypothesis is true' (personal communication with Jens R. Nyengaard). So we neither claim that we have a fitted model for minicolumns nor that the minicolumn hypothesis is true. 
Instead we have fitted a model with cylindrical clusters: from Table~\ref{t:MC_DLCPP} we see, if $|W_{xy}|$ denotes the area of $W_{xy}$, the estimated number of projected clusters is $|W_{xy}|\hat\kappa$, which is approximately $260$ for \texttt{L3} and $142$ for \texttt{L5}; the estimated expected size of a projected cluster is only 2.42 for \texttt{L3} and 3.87 for \texttt{L5}.
 \begin{table}[!ht]
	\centering
	\begin{tabular}{l|lll}
		& $\hat{\kappa}$ & $\hat{\sigma}$ & $\widehat{\alpha a}$ \\ \hline
		\texttt{L3} & 0.0040 & 5.45 & 2.42 \\
		\texttt{L5} & 0.0021 & 6.53 & 3.87 \\
	\end{tabular}
	\caption{
		For each dataset \texttt{L3} and \texttt{L5}, minimum contrast estimates for the parameters of our final model for $X_{xy}$ (the DLCPP model in Section~\ref{s:generalise_line_process}).
	}
	\label{t:MC_DLCPP}
\end{table}
Despite the expectation under the minicolumn hypothesis of having much higher values of $\widehat{\alpha a}$ than in Table~\ref{t:MC_DLCPP}, simulations of the fitted jinc-like DPP in the $xy$-plane seem in reasonable correspondence to the projected data; see Figure~\ref{f:pp_proj_and_sim_jDPP}. 
Furthermore, results from the GERL envelope procedure based on a concatenation of the $F$-, $G$-, and $J$-function do not provide any evidence against the jinc-like DPP model for the projected points with $p$-values of 0.67 for \texttt{L3} and 0.83 for \texttt{L5}.

\begin{figure}
	\centering
	\includegraphics[width=0.35\linewidth]{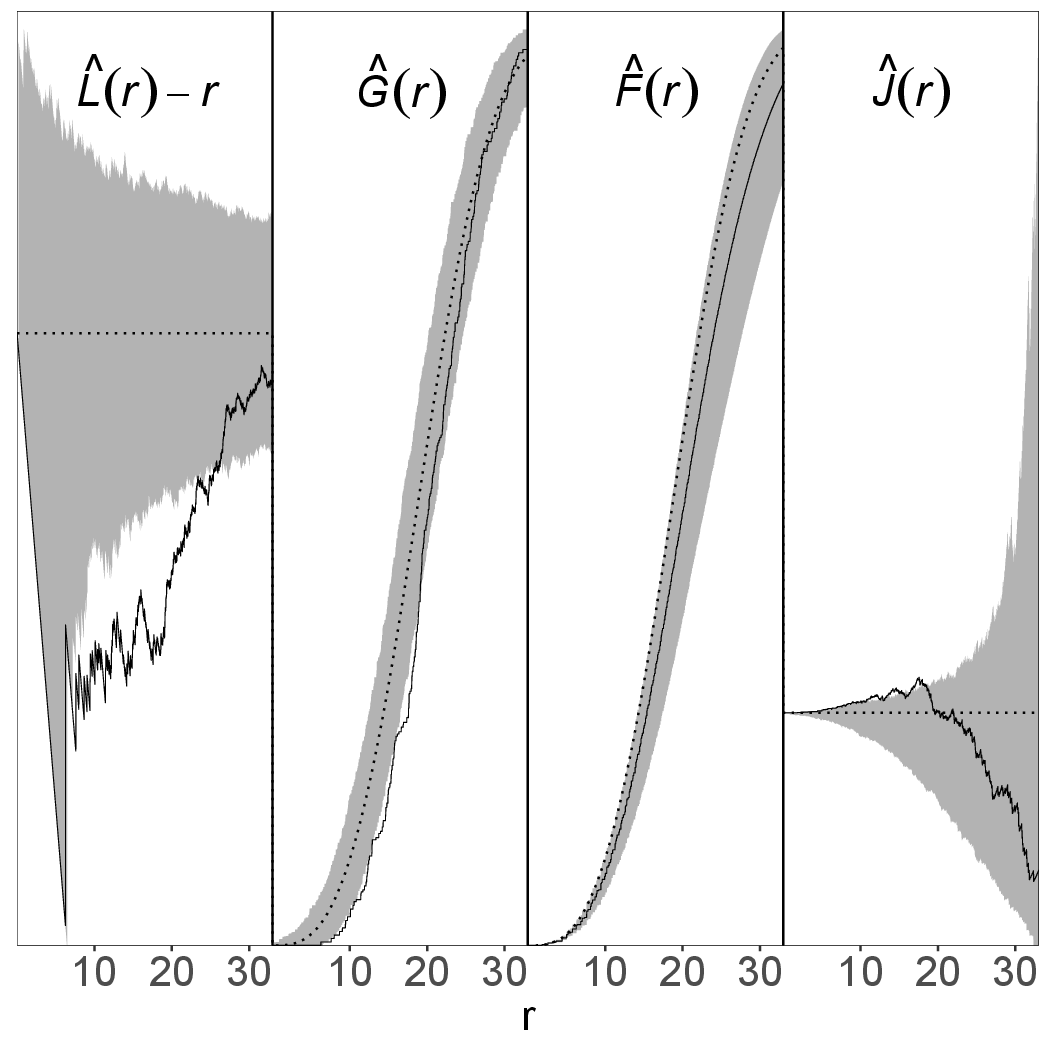}
	\includegraphics[width=0.64\linewidth, trim={14mm 52mm 6.5mm 60mm}, clip]{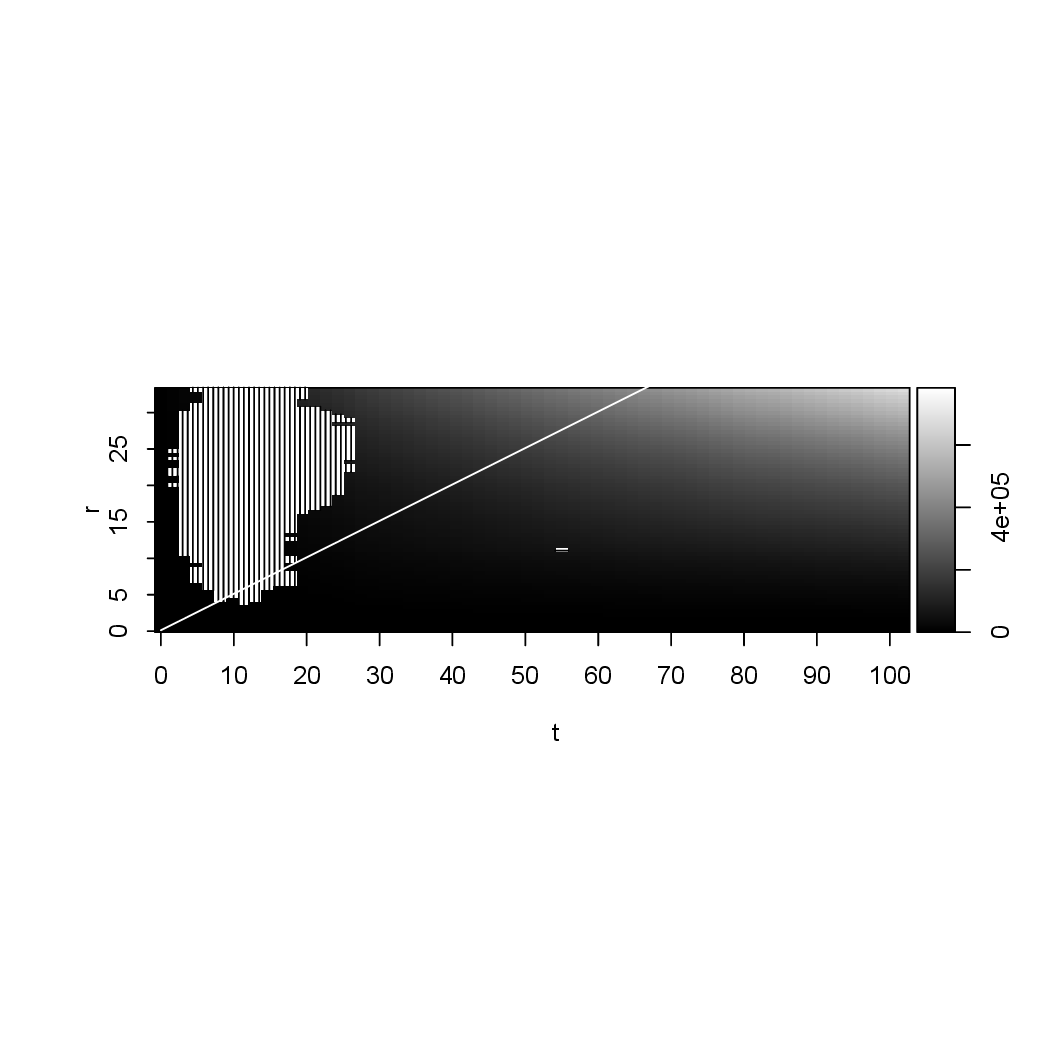}
	\includegraphics[width=0.35\linewidth]{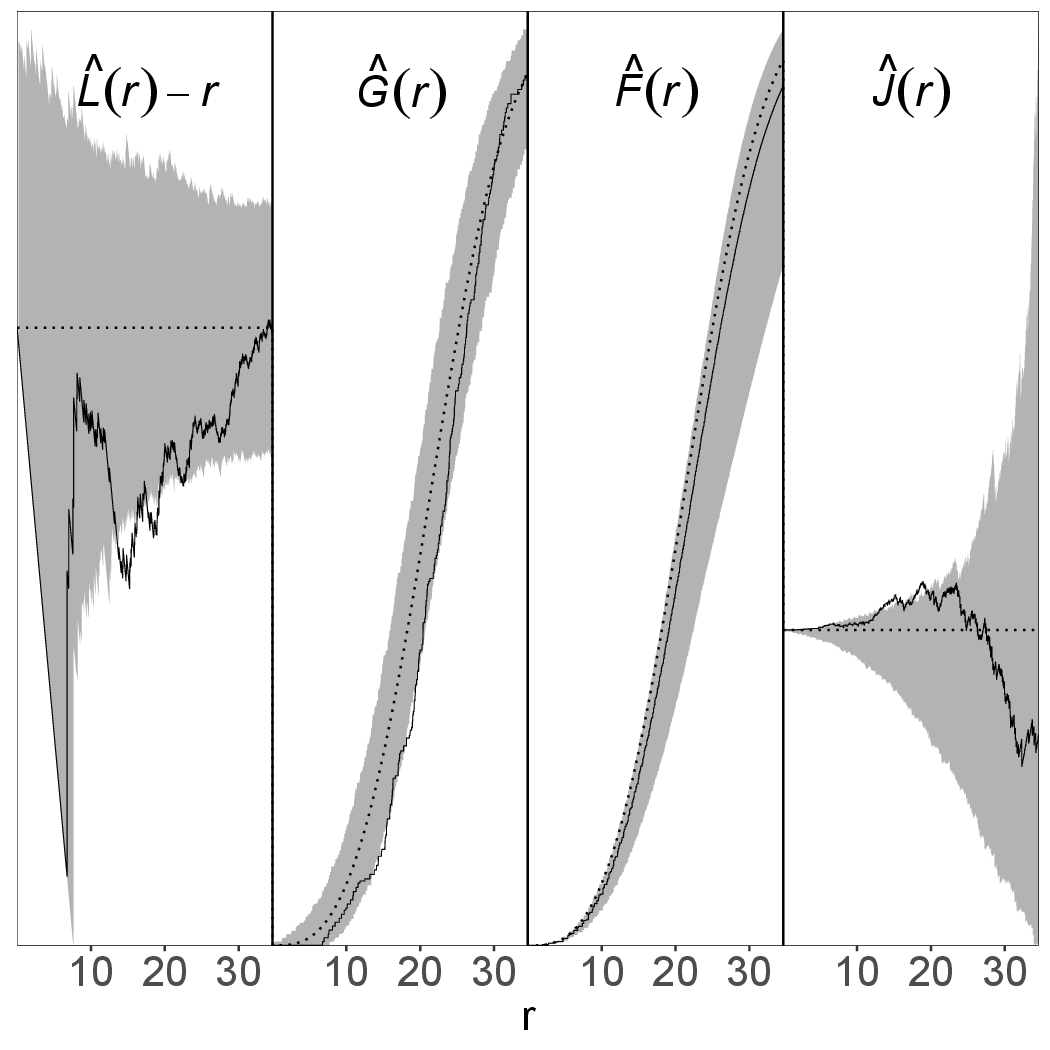}
	\includegraphics[width=0.64\linewidth, trim={14mm 52mm 6.5mm 60mm}, clip]{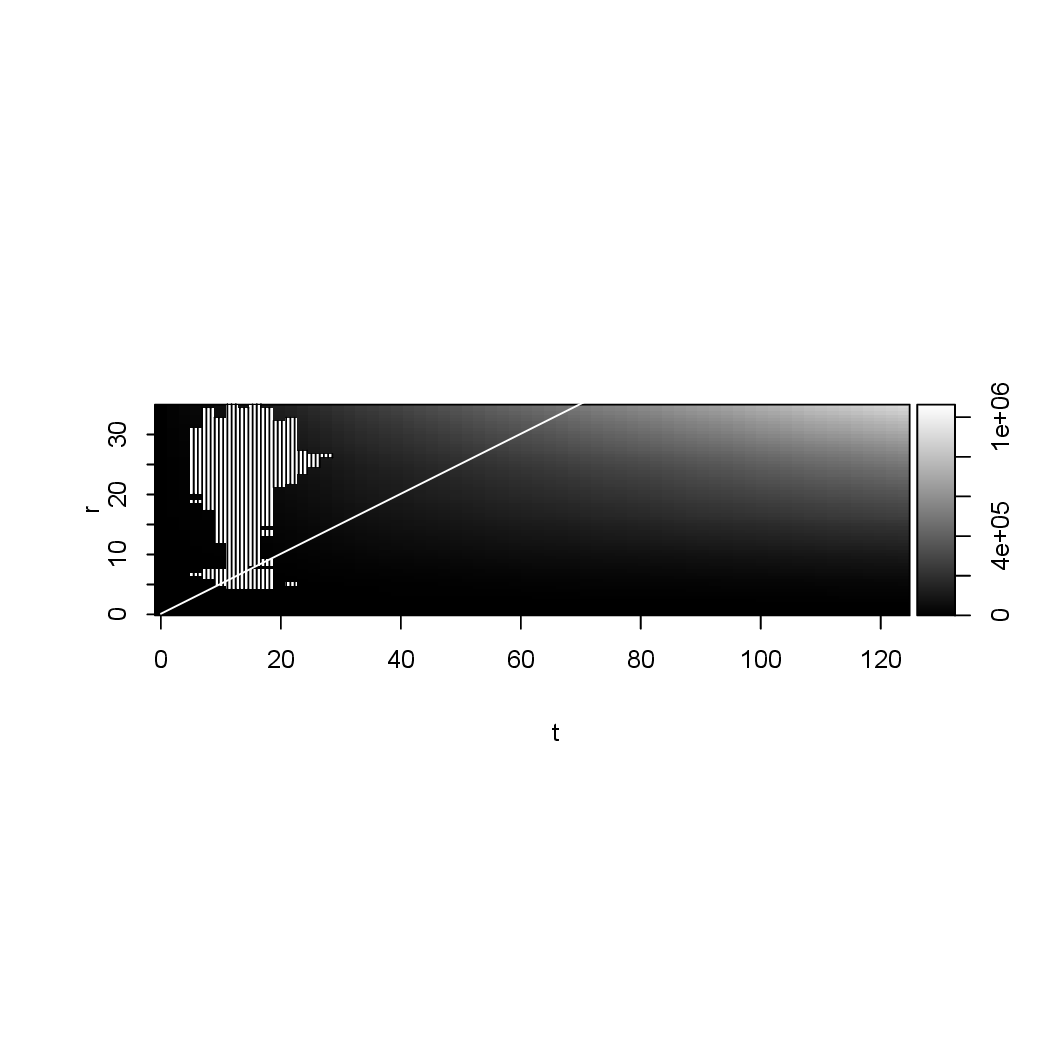}
	\caption{
		Results of the GERL envelope procedure under the fitted DLCPP. Left: concatenation of the one-dimensional empirical $L$-, $G$-, $F$-, and $J$-function for the data (solid line) together with $95\%$ GERL envelopes (grey region){, and theoretical function under CSR (dotted line)}; for ease of visualisation, the functions have been scaled. Right: empirical cylindrical $K$-function (grey scale) where shaded horizontal/vertical lines indicate that the function falls above/below the 95\% GERL envelope. The white line indicates the values for which the cylinder height is equal to the base diameter. Top: results for the dataset \texttt{L3}. Bottom: results for the dataset \texttt{L5}.
	}
	\label{f:summ_fun_DLCPP}
\end{figure}

Since the jinc-like DTPP model fits the projected data well, we proceeded and added independent uniform $z$-coordinates on $W_z$ to the simulations, thereby obtaining simulations of the jinc-like DLCPP. 
Figure~\ref{f:summ_fun_DLCPP} summarises the result of the 95\% GERL test based on the concatenation of {one-dimensional} functional summaries {as well as the cylindrical $K$-function} as described in Section~\ref{sec:def}. 
These plots show that the models do not account for the regularity of the data, but accounts for more clustering compared to the degenerate PLCPP discussed in Section~\ref{s:dPLCPP}.
This leads us to our next generalisation in Section~\ref{s:generalise_to_repulsiveness} {and the specific models in Section~\ref{sec:5.3}.}

\subsection{General MRF models for the $z$-coordinates given the $xy$-coordinates}\label{s:generalise_to_repulsiveness}\label{s:replacing}
Motivated by the observations at the end of the previous section, in this section we propose to model the vector of $z$-coordinates conditioned on the $xy$-coordinates 
by a pairwise interaction MRF given in \eqref{e:strauss_like} below. 

In general, conditioned on $X_{xy}=\{(x_i,y_i)\}_{i=1}^n$, we propose a model for $X_z$ with a conditional probability density function of the form
\begin{align}\label{e:strauss_like}
	f((z_i)_{i=1}^n \, |\, & (x_i,y_i)_{i=1}^n) \propto \gamma_1^{s_{B_1, \theta_1}((z_i)_{i=1}^n \, | \, (x_i,y_i)_{i=1}^n)}\gamma_2^{s_{B_2, \theta_2}((z_i)_{i=1}^n \, | \, (x_i,y_i)_{i=1}^n)}\\
	&\times\mathbb{I}(\|(x_i,y_i,z_i)-(x_j,y_j,z_j)\| > h \text{ for } 1 \le i < j \le n),\nonumber
\end{align}
with notation defined as follows. 
The right hand side in \eqref{e:strauss_like} is an unnormalised (conditional) density with respect to $n$-fold Lebesgue measure on $W_z$; $\mathbb{I}(\cdot)$ denotes the indicator function; and $\gamma_1>0$, $\gamma_2>0$, and $h \ge 0$ are unknown parameters. 
When $h>0$, it is a hard core parameter ensuring a minimum distance $h$ between all pair of points in $X$; for the pyramidal cell data it seems natural to include a hard core condition since the cells cannot overlap. Clearly, in order that the conditional density is well-defined, $h$ needs to be smaller that the length of $W_z$ (which is much larger than the diameter of a cell). 
Furthermore, for $k=1,2$,
\begin{equation*}
	s_{B_k,  \theta_k}((z_i)_{i=1}^n \, | \, (x_i,y_i)_{i=1}^n) = \sum_{1 \le i < j \le n} \mathbb{I}((x_i, y_i, z_i) \in  B_k(x_j, y_j, z_j; \theta_k)), 
\end{equation*}
where $B_k(x, y, z; \theta_k)\subset\mathbb R^3$ is an interaction region, with centre of mass $(x, y, z)$ and a `size and shape parameter', $\theta_k$, that determines the interaction between points. It is additionally assumed that the hard core ball, given by the three-dimensional closed ball of radius $h$ and centre $(x, y, z)$\, contains neither $B_1(x, y, z; \theta_1)$ nor $B_2(x, y, z; \theta_2)$. Finally, it is assumed that the symmetry condition \[\mbox{$(x_i,y_i,z_i)\in B_k(x_j, y_j, z_j; \theta_k)$ \quad if and only if\quad $(x_j,y_j,z_j)\in B_k(x_i, y_i, z_i; \theta_k)$}\] and the disjointness condition \[B_1(x,y,z; \theta_1) \cap B_2(x,y,z; \theta_2) = \emptyset\] are satisfied. The symmetry condition is imposed to ensure that we can interchange the roles of  $i$ and $j$.

If $\gamma_1 = \gamma_2 = 1$ and $h = 0$, then $X_z$ becomes the homogeneous binomial point process which depends only on $X_{xy}$ through the number of points in $X_{xy}$. In general, using Markov random field (MRF) terminology, $X_z$ conditioned on $X_{xy}=\{x_i,y_i\}_{i=1}^n$ is a pairwise interaction MRF with sites given by the lattice $\{x_i,y_i\}_{i=1}^n$, where two sites $(x_i,y_i)$ and $(x_j,y_j)$ are neighbours if and only if there exist $z_i',z_j'\in W_z$  and a $k\in\{1,2\}$ such that  $(x_i,y_i,z_i')\in B_k(x_j, y_j, z_j'; \theta_k)$ or $\|(x_i,y_i,z_i')-(x_j,y_j,z_j')\|\le h$. 
In other words, the conditional distribution of $z_i$ (the $i$'th coordinate in $X_z$) given both $\{x_i,y_i\}_{i=1}^n$ and all $z_j$ with $j\not=i$ depends only on those $z_j$ where $(x_i,y_i)$ and $(x_j,y_j)$ are neighbours.
We may also say that $(x_i, y_i, z_i)$ and $(x_j, y_j, z_j)$ interact if $(x_i, y_i, z_i)$ is in the union of $B_1(x_j, y_j, z_j; \theta_1)$ and $B_2(x_j, y_j, z_j; \theta_2)$ -- we refer to this union of sets as  the region of interaction of $(x_j, y_j, z_j)$ (note that we can interchange the roles of  $i$ and $j$) or if $\|(x_i,y_i,z_i)-(x_j,y_j,z_j)\|\le h$ (that is, the hard core condition is not satisfied, which happens with probability 0). The interaction can either cause repulsion/inhibition or attraction/clustering of the points in $X$ depending on whether $\gamma_k<1$ or $\gamma_k>1$ for $k=1,2$. 
Thus, apart from the hard core condition, the model allows for both repulsion and attraction but within different interaction regions $B_1$ and $B_2$.

Note that our hierarchical model construction yields a more flexible model for $X$ but we ignore edge effects in the sense that we have only specified a model for first $P_{xy}(Y_S)$ and second $X_z$ conditioned on $X_{xy}=P_{xy}(Y_S)\cap W_{xy}$, thereby ignoring a possible influence of points in $Y\setminus W$ when \eqref{e:strauss_like} is used in the latter step (unless it specifies  a binomial point process). 
This simplification is just made for mathematical convenience; indeed it would be interesting to construct a model taking edge effects into account so that $Y$ becomes stationary, but we leave this challenging issue for future research.

\subsection{Fitting specific MRF models for the $z$-coordinates given the $xy$-coordinates}\label{sec:5.3}
Below we first specify the ingredients of the conditional probability density function given in \eqref{e:strauss_like} for various models and discuss the overall conclusions, next describe how to find parameter estimates, and finally discuss how well the estimated models fit the data. 
Note that although we have not specified a stationary model for $Y$, it still make sense to interpret plots of the empirical cylindrical $K$-function and of the $\hat{F}$-, $\hat{G}$-, $\hat{J}$-, and $\hat L$-function, since we have stationarity in the $xy$-plane and approximately stationarity in the $z$-direction (as the density \eqref{e:strauss_like} is invariant under `translations of $(z_1,\ldots,z_n)$ within $W_z$').

\subsubsection{Selected models}
\begin{figure}
	\centering
	\begin{tikzpicture}[scale=.8]
		% thin cylinder bottom:
		\fill[left color=gray!50!black,right color=gray!50!black,middle color=gray!50,shading=axis,opacity=0.25] (1.1,-2.4) -- (1.1,0) arc (360:180:1.1cm and 0.275cm) -- (-1.1,-2.4) arc (180:360:1.1cm and 0.275cm);
		\draw (1.1,-0.42) -- (1.1,-2.4) arc (360:180:1.1cm and 0.275cm) -- (-1.1,-0.42); 
		\draw[densely dashed] (-1.1,0) arc (180:-180:1.1cm and 0.275cm);
		
		%%fat cylinder:
		%\draw[densely dashed] (-2,0) arc (180:0:2cm and 0.5cm);
		%\draw[densely dashed] (-4,2.3) arc (180:0:4cm and 1cm);
		
		%% hard ball:
		\fill[color=black, opacity = 1] (0,1.15) circle (0.625cm and 0.625cm);
		
		% fat cylinder:
		\fill[left color=gray!50!black,right color=gray!50!black,middle color=gray!50,shading=axis,opacity=0.5] (2,0) -- (2, 2.3) arc (360:180:2cm and 0.5cm) -- (-2,0) arc (180:360:2cm and 0.5cm);
		\fill[top color=gray!90!,bottom color=gray!2,middle color=gray!30,shading=axis,opacity=0.4] (0,2.3) circle (2cm and 0.5cm);
		\draw (-2,2.3) arc (180:123:2cm and 0.5cm);
		\draw (2,2.3) arc (0:57:2cm and 0.5cm);
		\draw (-2,2.3) arc (180:360:2cm and 0.5cm);
		\draw (-2, 2.3) -- (-2,0) arc (180:360:2cm and 0.5cm) -- (2, 2.3);
		
		% thin cylinder top:
		\draw[densely dashed] (-1.1,2.3) arc (180:0:1.1cm and 0.275cm);
		\fill[left color=gray!50!black,right color=gray!50!black,middle color=gray!50,shading=axis,opacity=0.25] (1.1,2.3) -- (1.1,4.7) arc (360:180:1.1cm and 0.275cm) -- (-1.1,2.3) arc (180:360:1.1cm and 0.275cm);
		\draw (1.1,4.7) -- (1.1,2.3) arc (360:180:1.1cm and 0.275cm) -- (-1.1,4.7) ++ (1.1,0) circle (1.1cm and 0.275cm);
		\fill[top color=gray!90!,bottom color=gray!2,middle color=gray!30,shading=axis,opacity=0.2] (0,4.7) circle (1.1cm and 0.275cm);
		
		% names on cylinder shapes
		\node at (-2.65,1.15) {$B_1$};
		\node at (4.2,1.15) {$B_2$};
		\draw (4,1.65) -- (1.5,4);
		\draw (4,0.65) -- (1.5,-1.85);
	\end{tikzpicture}
	\caption{Visualisation of the hard core region ball (in dark) and the cylindrical interaction regions $B_1$ (the cylinder) and $B_2$ (the union of the two elongated cylinders) used in our final model 5 for \texttt{L3}. The relative dimensions of the objects correspond to the parameter estimates in Table~\ref{t:final}.}
	\label{f:interactionregion}
\end{figure}
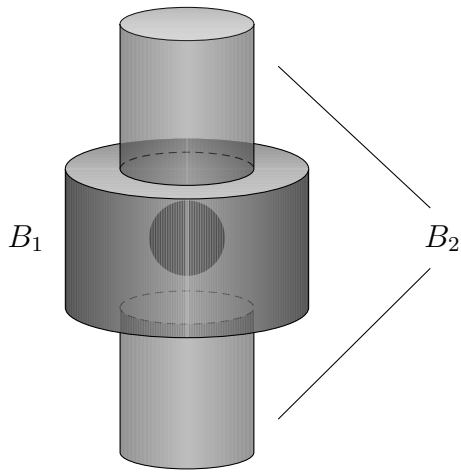

In our search for a suitable model for the nucleolus locations, we considered many special cases of \eqref{e:strauss_like}. Table~\ref{t:z_models} summarises five selected models, where $b((x,y,z); r)$ is the ball with centre $(x,y,z)$ and radius $r$, and where $c((x,y,z); r, t)$ and $d((x,y,z); r, t)$ denote the cylinder and double cone, respectively, with centre of mass at $(x,y,z)$, height $2t$, base radius $r$, and extending in the $z$-direction. Note that in Table~\ref{t:z_models} we do not need to specify $B_k(\cdot; \theta_k)$ when $\gamma_k = 1$. 
For the final model 5, \eqref{e:strauss_like} becomes the conditional density  
\begin{align*}
	f((z_i)_{i = 1}^n\, |\, (x_i, y_i)_{i = 1}^n)
	\propto  \prod_{1 \leq i < j \leq n} &\mathbb{I}(\|(x_i,y_i,z_i)-(x_j,y_j,z_j)\| > h)\\
	\times&\gamma_1^{\mathbb{I}( \| (x_i, y_i) - (x_j, y_j)\| \leq r_1, \,  |z_i - z_j| \leq t_1)}\\
	\times& \gamma_2^{\mathbb{I}( \| (x_i, y_i) - (x_j, y_j)\| \leq r_2, \, t_1 < |z_i - z_j| \leq t_2)},
\end{align*}
where the cylindrical interaction regions $B_1(x_j,y_j,z_j;\theta_1)=c(x_j,y_j,z_j;r_1,t_1)$ and $B_2(x_j,y_j,z_j;\theta_2)=c(x_j,y_j,z_j;r_2,t_2)\setminus c(x_j,y_j,z_j;r_1,t_1)$ are illustrated in Figure~\ref{f:interactionregion}, $h \ge 0$ is still the hard core parameter, $\gamma_1>0$ and $\gamma_2 >0$ are interaction parameters, and $0< r_2 \leq r_1$ and $0 < t_1 < t_2$ are parameters which determine the `range of interaction' {satisfying} $h < \sqrt{t_k^2 + r_k^2}$ for $k = 1, 2$. These restrictions on the parameters are empirically motivated by use of functional summaries as detailed below.

First, we considered model~1 which is a hard core model if $h>0$ and one of the simplest ways of modelling regularity; 
note that model~1 with $h=0$ is the binomial point process with a uniform density as considered in Section~\ref{s:dPLCPP}.
Though accounting for small distance repulsion, when fitted to the data, model~1 turned out not to account for the repulsion at larger scales. Second, we considered model~2 which is a conditional Strauss model with a hard core condition \citep[see][and the references therein]{MW:2004}. 
For this model the scale of repulsion for the $z$-coordinates seemed too great for points with similar $xy$-coordinates, and therefore we found it natural to replace the spherical interaction region with a cylinder, yielding model~3.
However, model~3 did not correct the problem, and continuing with a single region of interaction we next suggested model~4 with a region given by a cylinder minus a double cone. Model~4 does to a smaller degree penalise the occurrence of points with similar $xy$-coordinates. However, this model was not suitable either. Models~1--4 were discarded by GERL tests with extremely small $p$-values. 
Finally, we considered model~5 which is a more flexible model that allows for both repulsion and aggregation within cylinder shaped interaction regions, cf.\ the discussion in {Section~\ref{s:model2}}.
For simplicity all the models were also considered without a hard core condition, that is $h=0$, but was in every case found inadequate. 
\begin{table}[!ht]
	\centering
	\begin{tabular}{c|ccccc}
		Model & $\gamma_1$ & $\gamma_2$ & $B_1(\cdot;\theta_1)$ & $B_2(\cdot;\theta_2)$ & $\theta_1$ \\ \hline
		1 & 1 & 1 & - & - & - \\
		2 & $>0$ & 1 & $b(\cdot; r)$ & - & $r>h$ \\
		3 & $>0$ & 1 & $c(\cdot; r, t)$ & - & $r,t>0$ \\
		4 & $>0$ & 1 & $c(\cdot; r, t)\backslash d(\cdot; r, t)$ & - & $r,t>0$ \\
		5 & $>0$ & $>0$ & $c(\cdot; r_1, t_1)$ & $c(\cdot; r_2, t_2) \setminus c(\cdot; r_1, t_1)$ & $r_1, t_1>0$  \\
	\end{tabular}
	\caption{Specific choices of the parameters $\gamma_1,\gamma_2,\theta_1,\theta_2$ and the interaction regions $B_1(\cdot;\theta_1),B_2(\cdot;\theta_2)$ for five models given by the density \eqref{e:strauss_like}. For each model, a hard core parameter $h\ge0$ is included. Apart from the specified restrictions, it is required for models~2--5 that $B_1(\cdot;\theta_1)\not\subseteq b(\cdot;h)$ (for model~2 this means that $r>h$ as already indicated) and in addition for model~5 that $B_2(\cdot;\theta_2)\not\subseteq b(\cdot;h)$ where $\theta_2=(r_2,t_2)$ with $r_1 \geq r_2>0$ and $t_2> t_1$.}
	\label{t:z_models}
\end{table}

\subsubsection{Results based on maximum pseudo likelihood}\label{sec:MPLE}
The likelihood function corresponding to \eqref{e:strauss_like} involves a normalising constant which needs to be approximated by Markov chain Monte Carlo methods. We propose an easier alternative based on the pseudo likelihood function \citep{besag:75} defined as follows when the data is given by $\{(x_i,y_i,z_i)\}_{i=1}^n\subset W$. For $i=1,\ldots,n$,
the $i$'th full conditional density associated to \eqref{e:strauss_like} is

\begin{align}
	f(z_i\,|\, & (z_1,\ldots,z_{i-1},z_{i+1},\ldots,z_n),(x_j,y_j)_{j=1}^n) \nonumber\\ 
	 = &\,{\mathbb{I}(\|(x_i,y_i,z_i)-(x_j,y_j,z_j)\| > h \text{ for } j \ne i)\gamma_1^{s_{1,i}}\gamma_2^{s_{2,i}}}/c_i
	 %{c_i(\gamma_1, \gamma_2, h, \theta_1, \theta_2\,|\,\{z_1,\ldots,z_{i-1},z_{i+1},\ldots,z_n\},\{(x_j,y_j)\}_{j=1}^n)} 
	 \label{e:marginal_density}
\end{align}
where we define
\[s_{k,i}=\sum_{j:\, j\ne i}\mathbb{I}\left((x_j, y_j, z_j)\in B_k((x_i, y_i, z_i);\theta_k)\right),\qquad k=1,2,\]		
and where the normalising constant is given by
\begin{align*} 
	c_i&= \sum_{k=0}^{n-1}\sum_{l=0}^{n-1} \gamma_1^k \gamma_2^l
	\int_{W_z} \mathbb{I}(\|(x_i,y_i,z)-(x_j,y_j,z_j)\| > h \text{ for } j \ne i) \\
	&\qquad\qquad\qquad\qquad \times\mathbb{I}\bigg(\sum_{j:\, j\ne i}\mathbb{I}\left((x_j, y_j, z_j)\in B_1((x_i, y_i, z);\theta_1)\right) = k\bigg) \\
	& \qquad\qquad\qquad\qquad\times\mathbb{I}\bigg(\sum_{j:\, j\ne i}\mathbb{I}\left((x_j, y_j, z_j)\in B_2((x_i, y_i, z);\theta_2)\right) = l\bigg)
	\, \mathrm{d} z.
\end{align*}

To estimate the model parameters we maximise the log pseudo likelihood given by 
\begin{equation}
	\begin{aligned}\label{e:pseudo_likelihood}
		LP(\gamma_1, &\gamma_2, h, \theta_1, \theta_2)\\
		&= \sum_{i = 1}^{n}\log f(z_i\,|\, (z_1,\ldots,z_{i-1},z_{i+1},\ldots,z_n),(x_j,y_j)_{j=1}^n).
	\end{aligned}
\end{equation}

Clearly, by \eqref{e:marginal_density} the maximum pseudo likelihood estimate (MPLE) $\hat{h}$ of $h$ is the minimum distance between any distinct pair of points $(x_i,y_i,z_i)$ and $(x_j,y_j,z_j)$ in the data. This in fact also corresponds to the maximum likelihood estimate.
For $h = \hat{h}$ and for fixed $\theta_1$ and $\theta_2$, we easily obtain the following. For each of models~2--4, the MPLE of $\gamma_1$ exists if and only if $s_{1,i}\not=0$ for some $i$, and then the log pseudo likelihood function is strictly concave with respect to $\log\gamma_1$. For model~5, the MPLE of $(\gamma_1,\gamma_2)$ exists if and only if $s_{1,i}\not=0$ for some $i$ and $s_{2,j}\not=0$ for some $j$, and then the log pseudo likelihood function is strictly concave with respect to $(\log\gamma_1,\log\gamma_2)$. 
Therefore, the (profile) log pseudo likelihood can be maximised by a combination of a grid search over $\theta_1$ and $\theta_2$ and numerical optimisation with respect to $\gamma_1$ and $\gamma_2$. 

Each of the five models in Table~\ref{t:z_models} were fitted to \texttt{L3} and \texttt{L5} by finding the maximum pseudo likelihood estimate, where for
the numerical optimisation we used 
\texttt{optim} (a general-purpose optimisation function from the \texttt{R}-package \texttt{stats}). Table~\ref{t:final} shows the maximum pseudo likelihood estimates of model~5 for the two datasets. 
Most notably, around each point in $X$, there is a repulsive interaction region (since $\hat{\gamma}_1<1$) within a stunted cylinder $B_1$ and an attractive interaction region (since $\hat{\gamma}_2>1$) within the union $B_2$ of two elongated cylinders, see again Figure~\ref{f:interactionregion}. In particular the fitted model is in accordance to the empirical findings as noted later when the cylindrical $K$-function {extending in the $z$-direction} of Figure~\ref{f:summ_fun_CSR} is discussed.
Specifically, there is repulsion between two points $(x_i,y_i,z_i)$ and $(x_j, y_j, z_j)$ in $X$  if $(x_i, y_i)$ and $(x_j, y_j)$ lie within distance 20\,$\mu\mathrm{m}$ for \texttt{L3} and 24.25\,$\mu\mathrm{m}$ for \texttt{L5} and if $z_i$ and $z_j$ lie within distance 11.5\,$\mu\mathrm{m}$ for \texttt{L3} and 15.5\,$\mu\mathrm{m}$ for \texttt{L5}. Analogously, there is attraction between $(x_i,y_i,z_i)$ and $(x_j, y_j, z_j)$ if $(x_i, y_i)$ and $(x_j, y_j)$ lie within distance 11\,$\mu\mathrm{m}$ for \texttt{L3} and 14.75\,$\mu\mathrm{m}$ for \texttt{L5} and if $|z_i - z_j|$ is between 11.5--35.5\,$\mu\mathrm{m}$ for \texttt{L3} and 15.5--37.25\,$\mu\mathrm{m}$ for \texttt{L5}. Moreover, the estimated hard core $\hat h$ is between 6-7\,$\mu\mathrm{m}$, which is in accordance with `distance between the nucleolus and the membrane of a pyramidal cell' (personal communication with Jens R.\ Nyengaard). 
Note that $2\hat h$ (the diameter  of an estimated hard core ball) is about half as small as $2\hat t_1$ (the estimated height of $B_1$).
Finally, comparing Tables~\ref{t:MC_DLCPP}-\ref{t:final}, we note that the two `clustering parameters'  $2\hat \sigma$ and $\hat{r}_2$ are of the same order.
\begin{table}[!ht]
	\centering
	\begin{tabular}{c|lllllll}
		& $\hat{\gamma_1}$ & $\hat{\gamma_2}$ & $\hat{h}$ & $\hat{r}_1$ & $\hat{t}_1$ & $\hat{r}_2$ & $\hat{t}_2$ \\ \hline
		\texttt{L3} & 0.41 & 1.78 & 6.25 & 20 & 11.5 & 11 & 35.5 \\
		\texttt{L5} & 0.51 & 1.68 & 6.77 & 24.25 & 15.5 & 14.75 & 37.25 \\
	\end{tabular}
	\caption{
		Pseudo likelihood estimates of our final model (model~5 from Table~\ref{t:z_models} in Section~\ref{s:generalise_to_repulsiveness}) for the datasets \texttt{L3} and \texttt{L5}.
	}
	\label{t:final}
\end{table}

\begin{figure}
	\centering
	\includegraphics[width=0.32\linewidth]{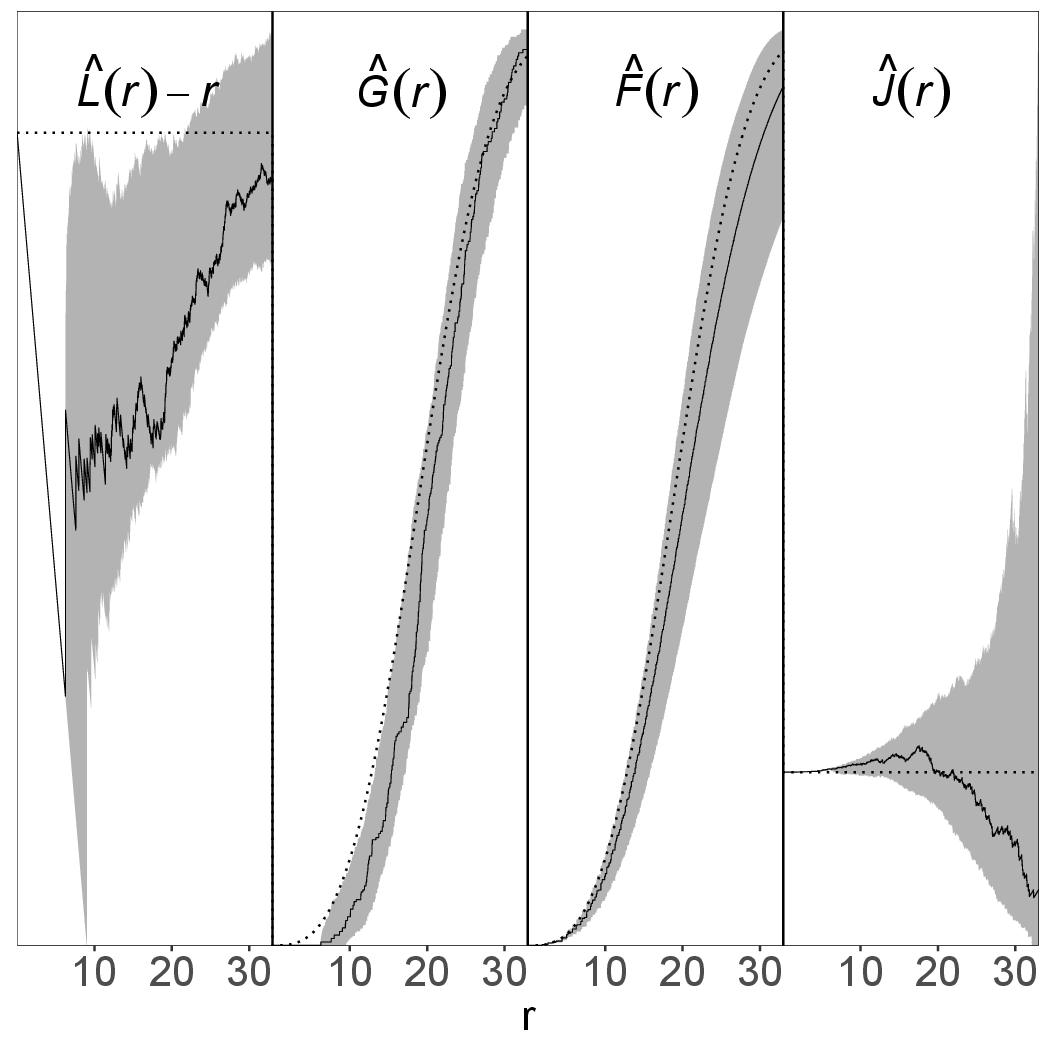}
	\includegraphics[width=0.64\linewidth, trim={14mm 52mm 6.5mm 60mm}, clip]{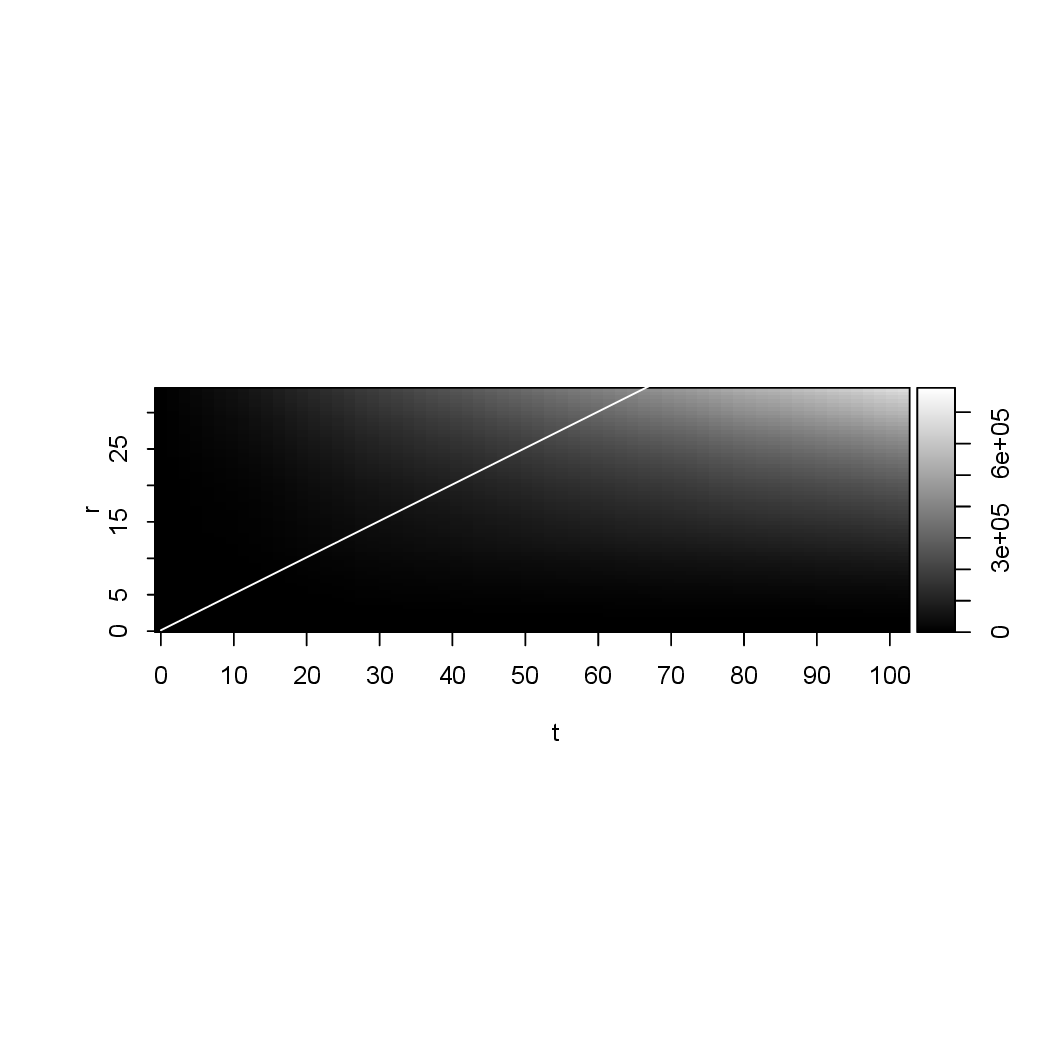}
	\includegraphics[width=0.32\linewidth]{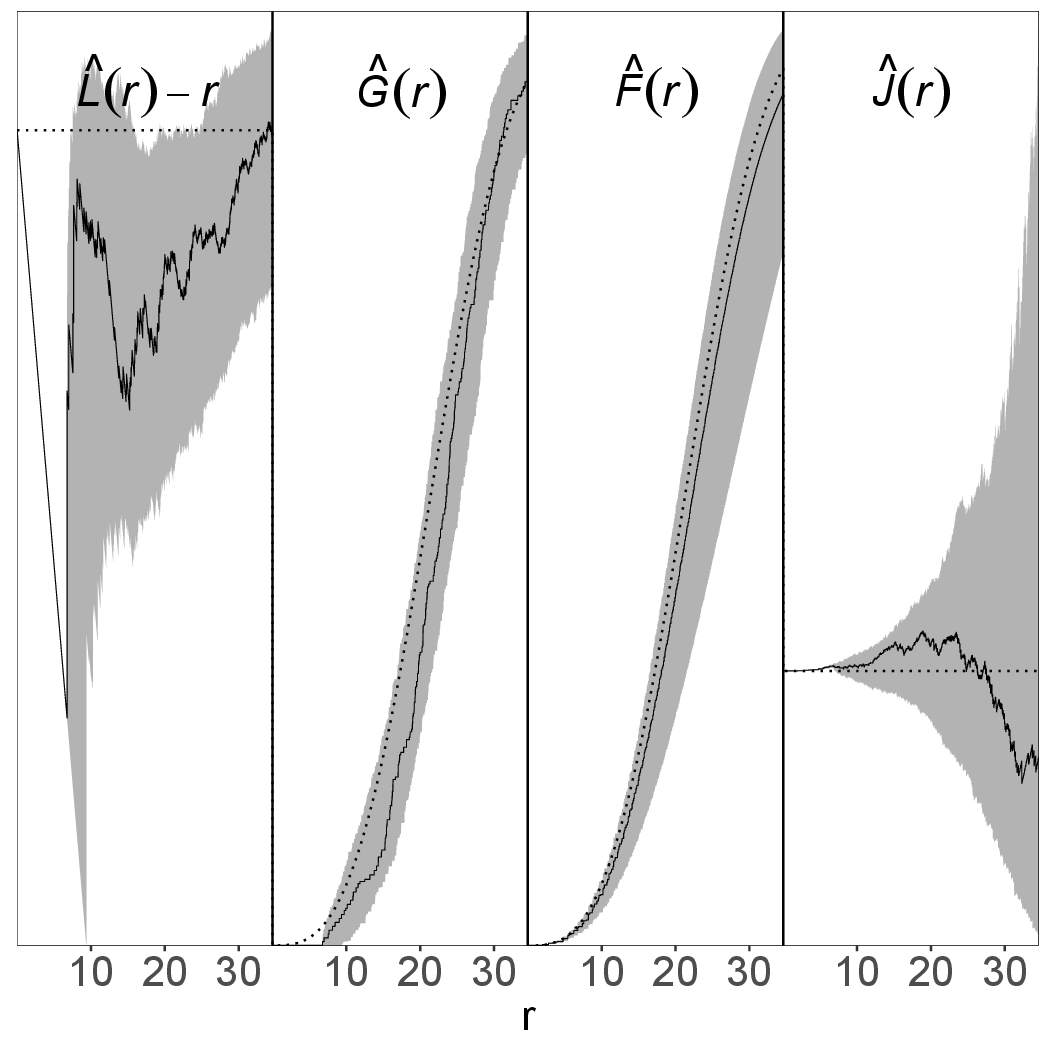}
	\includegraphics[width=0.64\linewidth, trim={14mm 52mm 6.5mm 60mm}, clip]{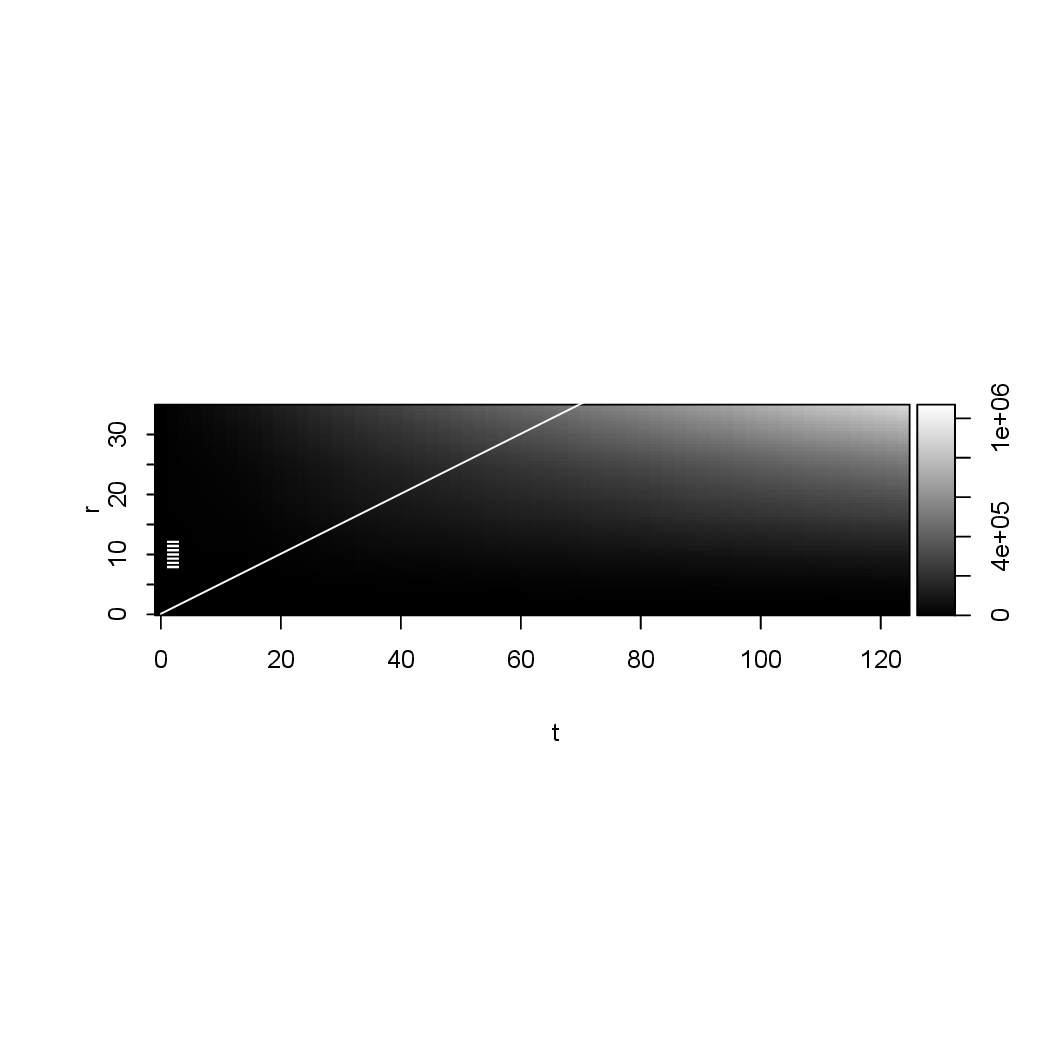}
	\caption{
		Results of the GERL envelope procedure under the fitted model~5. Left: concatenation of the one-dimensional empirical $L$-, $G$-, $F$-, and $J$-function for the data (solid line) together with $95\%$ GERL envelopes (grey region){, and theoretical function under CSR (dotted line)}; for ease of visualisation, the functions have been scaled. Right: empirical cylindrical $K$-function (grey scale) where shaded horizontal/vertical lines indicate that the function falls above/below the 95\% GERL envelope. The white line indicates the values for which the cylinder height is equal to the base diameter. Top: results for the dataset \texttt{L3}. Bottom: results for the dataset \texttt{L5}.
	}
	\label{f:summ_final}
\end{figure}

Model checking was performed using GERL envelope procedures based on the concatenation of {one-dimensional} functional summaries as well as the cylindrical $K$-function as discussed in Section~\ref{sec:GERL}. For the fitted models, model~5 was the most appropriate with $p$-values of 0.15 and 0.32 for \texttt{L3} and 0.23 and 0.02 for \texttt{L5} when using the GERL envelope procedure based on the concatenation of the one-dimensional summary functions ($L-r$, $G$, $F$, $J$) and the cylindrical $K$-function in the $z$-direction, respectively; the 95\% GERL envelopes are visualised in Figure~\ref{f:summ_final}.
Thus no evidence is seen against the fitted models summarised in Table~\ref{t:final} for \texttt{L3} while only slight evidence is present for \texttt{L5}.

Finally, note that simulations from each of models~1--5 can straightforwardly be obtained using 
a Metropolis-Hastings algorithm for a fixed number of points and given a realisation of the $xy$-coordinates.
Specifically, we used Algorithm~7.1 in \cite{MW:2004} but with a systematic updating scheme cycling over the point indexes 1 to $n$, using a uniform proposal for a new point in $W_z$ and a Hastings ratio calculated from the full conditional \eqref{e:marginal_density}.
We successively updated each point 100 times under the systematic updating scheme, corresponding to 63400 and 54800 point updates for $\texttt{L3}$ and $\texttt{L5}$,  respectively.

\section{Concluding remarks}\label{s:discussion}
The structure of minicolumns (clusters) of points has been investigated in relation to illnesses such as Down’s
syndrom \citep{Buxhoeveden:etal:2002}, schizophrenia \citep{Casanova:2007}, autism \citep{Casanova:etal:2006}, and Alzheimer’s disease \citep{Esiri:Chance:2006,Chance:etal:2011}. However, none of the methods used treat the positions of cells as a 3D point pattern. 
This paper has contributed to 3D point process methodology by developing new models and inference procedures based on a combination of moment based estimation and maximum pseudo likelihood estimation, and by illustrating the usefulness of various 3D functional summaries, in particular the cylindrical $K$-function, in combination with global rank envelope test procedures.  

We fitted several hierarchical models for the two 3D point pattern datasets of nucleolus locations and related our findings to the minicolumn hypothesis, which claims that minicolumns  extend parallel to the $z$-axis. Starting with a homogeneous Poisson process and second with the degenerated Poisson line cluster point process model in \cite{Moller:etal:2016}, we proceeded by developing more advanced point process models for a columnar structure along the $z$-axis. 
Our final model can be summarized as follows. For the $xy$-coordinates, we argued a need for a cluster point process given by a generalized shot noise Cox process, where the cluster centres are described by a repulsive point process given by a stationary determinantal point process, and where the offspring distribution is given by a bivariate isotropic normal distribution. For the $z$-coordinates conditioned on the $xy$-coordinates, we specified a pairwise interaction Markov random field model with repulsion between nucleolus locations given by a hard core condition on a small scale and a stunted cylindrical interaction region on a larger scale, as well as clustering between nucleolus locations given by an elongated cylindrical interaction region. The final model specifies  much smaller columns than expected under the minicolumn hypothesis.
Although the same type of model describes the two datasets, very different parameter estimates were obtained. 

All the studies in \cite{Buxhoeveden:etal:2002}, \cite{Casanova:etal:2006}, \cite{Casanova:2007}, and \cite{Chance:etal:2011} show deviations in the minicolumn structure in the brains of sick subjects compared to normal ones. For future applications with several 3D point pattern datasets belonging to different groups (related to normal and sick objects),  it remains to develop statistical methods for comparison of the groups. We hope that our methodology may serve as an inspiration for this and other future developments.

\section*{Acknowledgements}
This work was supported by The Danish Council for Independent Research | Natural Sciences, grant DFF -- 7014-00074
`Statistics for point processes in space and beyond', and by the `Centre for Stochastic Geometry and
Advanced Bioimaging', funded by grant 8721 from the Villum Foundation. We are thankful to Ali H.\ Rafati for collecting the data analysed in this paper and to Jens R.\ Nyengaard and Ninna Vihrs for helpful comments.

\appendix
\counterwithin{figure}{section}
\counterwithin{table}{section}
\section{Appendix} \label{app:sim_stud}
A simulation study was performed in order to investigate the performance of the maximum pseudo likelihood estimation procedure used for fitting our final model (model 5 from Table~\ref{t:z_models}) with a DTPP for the centre points in the $xy$-plane. The simulations were obtained by
\begin{enumerate}
	\item simulating 100 DTPPs with $\kappa=0.0040$, $\sigma=5.45$, and $\alpha a=2.42$, corresponding to the parameter estimates in Table~\ref{t:MC_DLCPP} for the dataset \texttt{L3};
	\item for each of the 100 simulated DTPPs, simulating the associated vector of $z$-coordinates from the pairwise interaction MRF specified by the conditional density \eqref{e:strauss_like} and with regions of interaction as specified for model 5 in Table~\ref{t:z_models}. For this we used a Metropolis-Hastings algorithm as described in Section~\ref{s:replacing} with $\gamma_1=0.41$, $\gamma_2=1.78$, $h=6.25$, $r_1=20$, $t_1=11.5$, $r_2=11$, $t_2=35.5$, corresponding to the parameter estimates in Table~\ref{t:final} for the dataset \texttt{L3}.
\end{enumerate}

\begin{table}[!ht]
	\centering
	\begin{tabular}{l|lllllll}
		& ${\gamma_1}$ & ${\gamma_2}$ & ${h}$ & ${r}_1$ & ${t}_1$ & ${r}_2$ & ${t}_2$ \\ \hline
		True value         & 0.41  & 1.78  & 6.25  & 20    & 11.5   & 11    & 35.5  \\
		Mean               & 0.40  & 1.82  & 6.73  & 19.96 & 11.56  & 11.06 & 35.18 \\
		Standard deviation & 0.048 & 0.081 & 0.447 & 0.608 & 0.292  & 0.618 & 1.357 \\
	\end{tabular}
	\caption{Parameter values, their estimate and standard deviation from the simulation study.}
	\label{t:simstud_parms}
\end{table}

\begin{figure}
	\centering
	\includegraphics[width=0.25\linewidth]{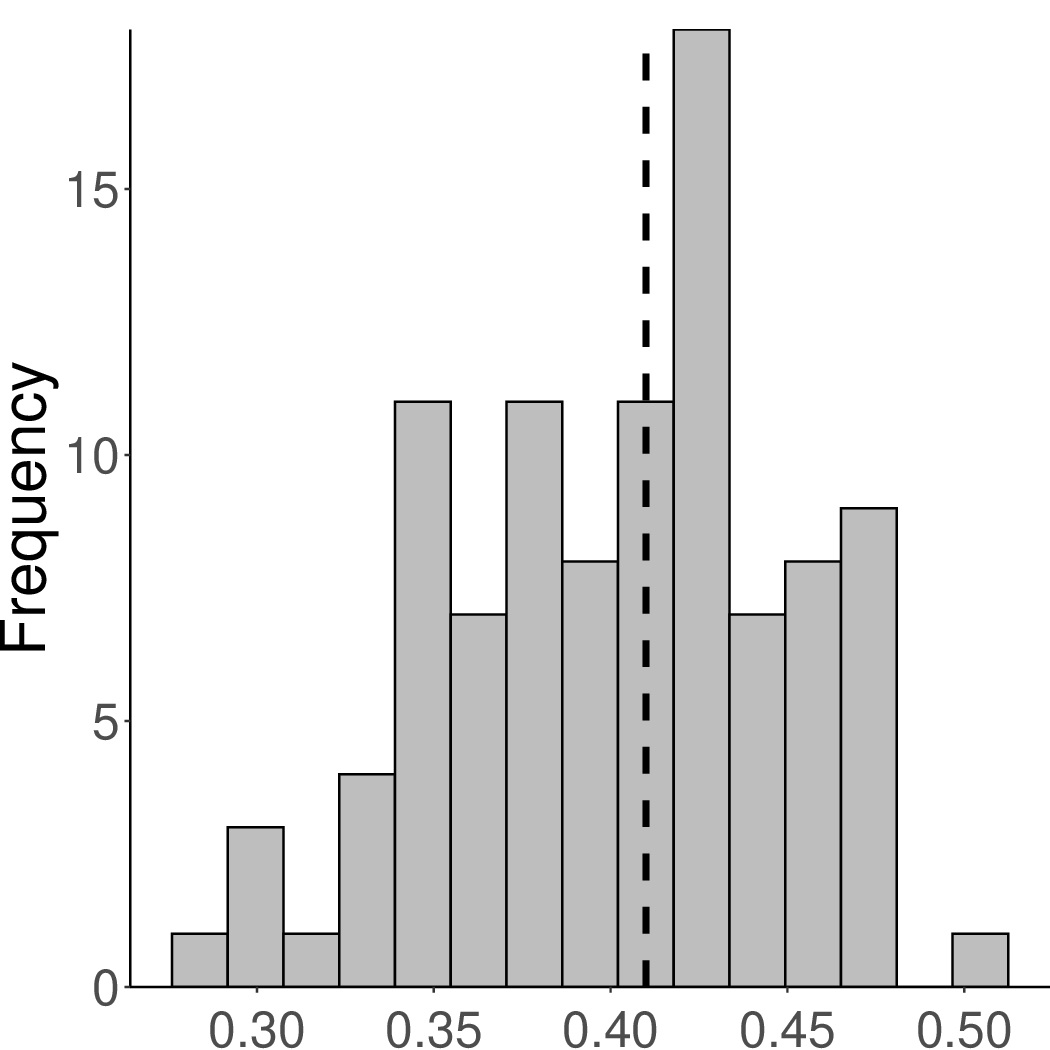}%
	\includegraphics[width=0.25\linewidth]{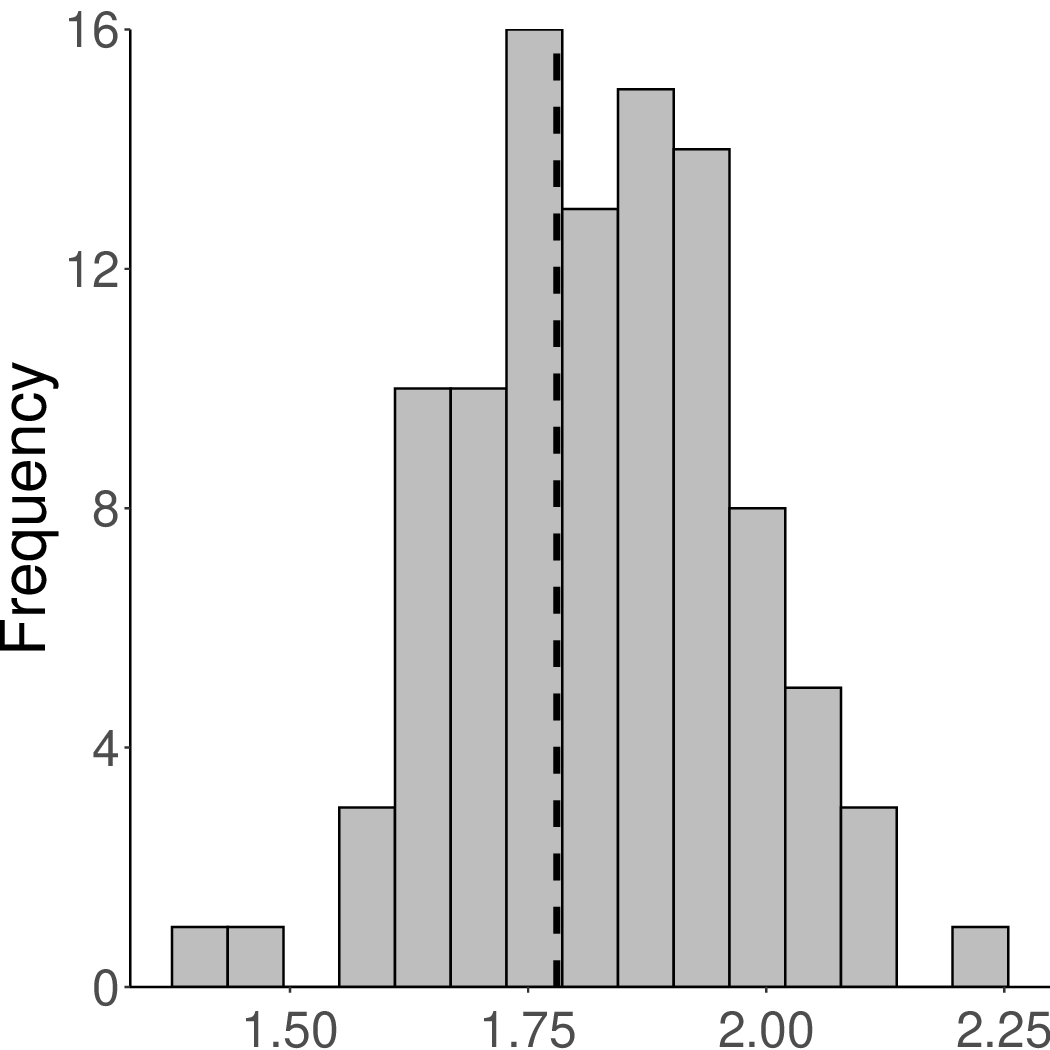}%
	\includegraphics[width=0.25\linewidth]{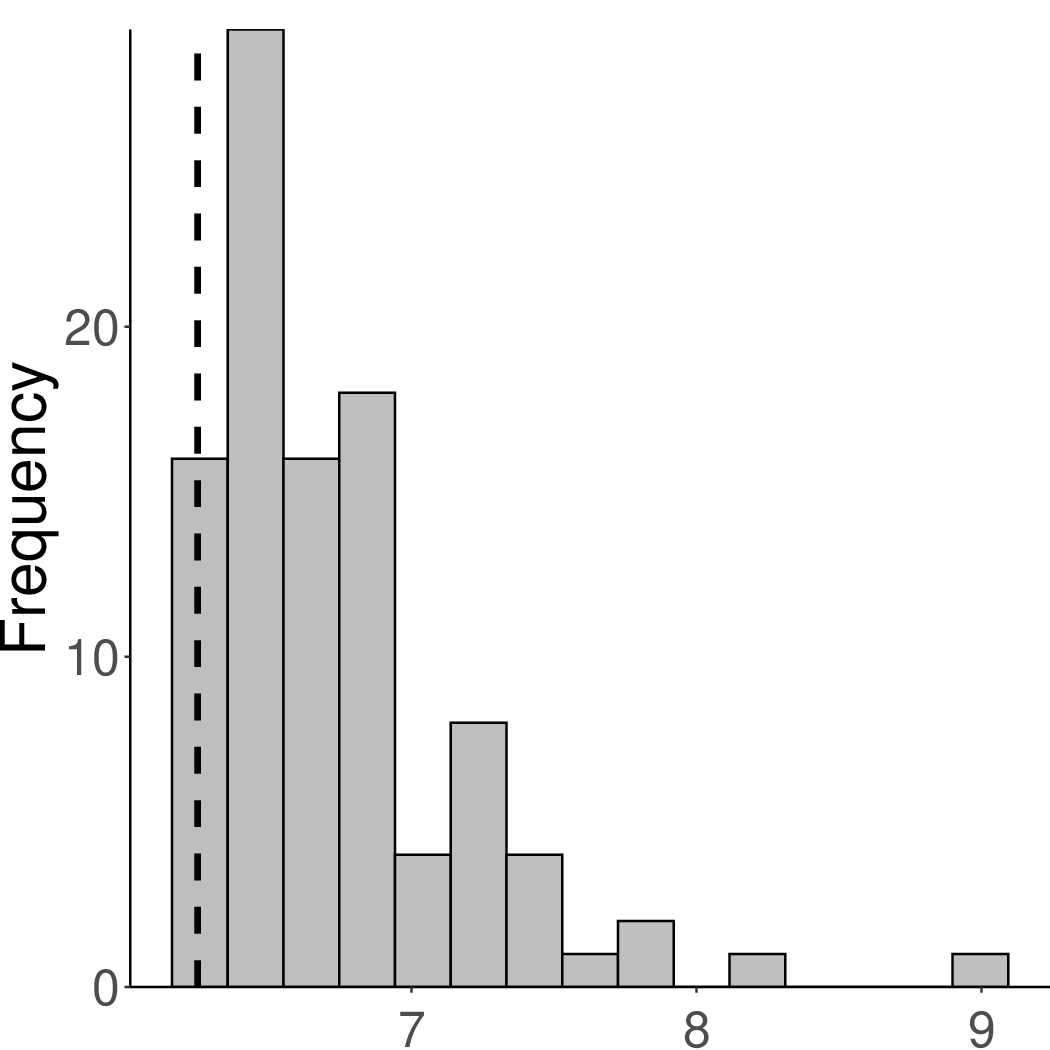}%
	\includegraphics[width=0.25\linewidth]{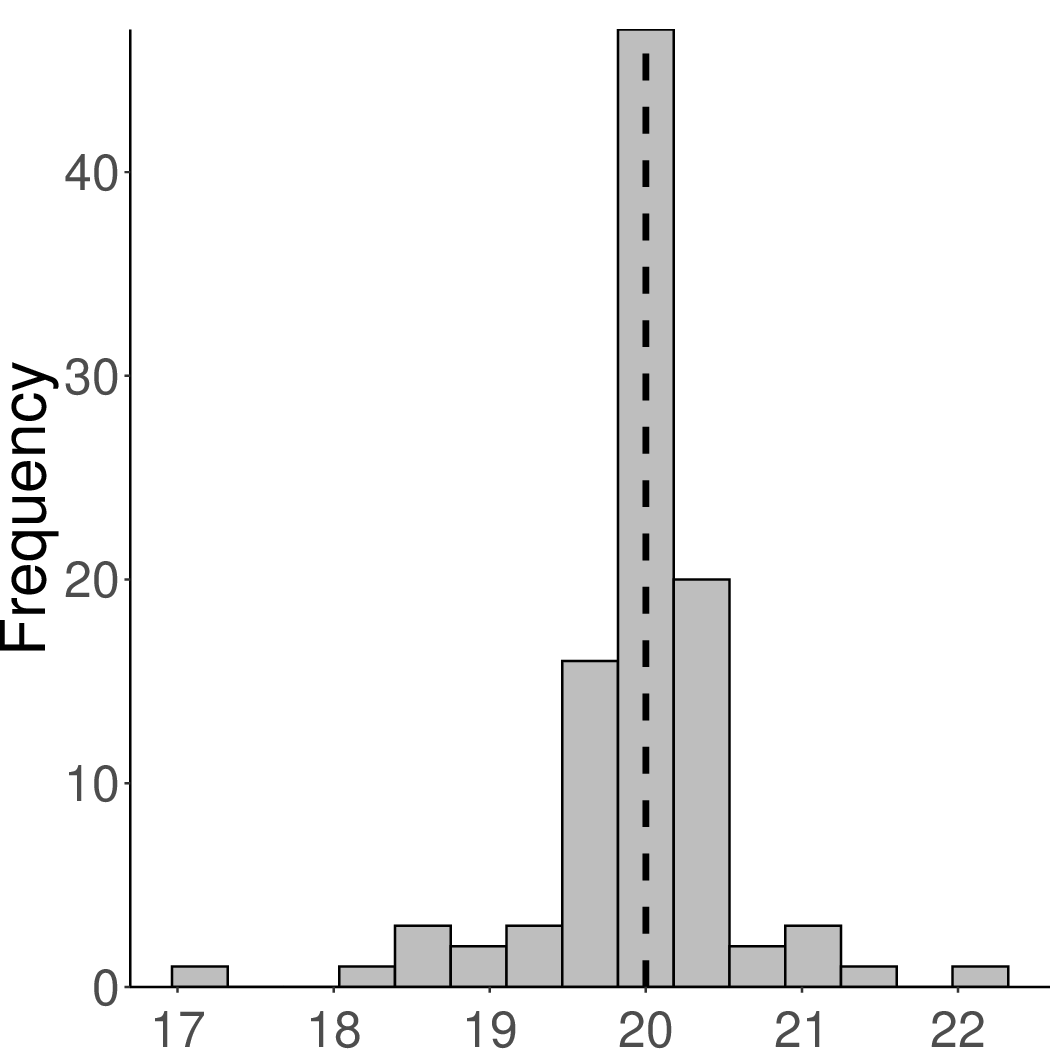}%
	\\
	\includegraphics[width=0.25\linewidth]{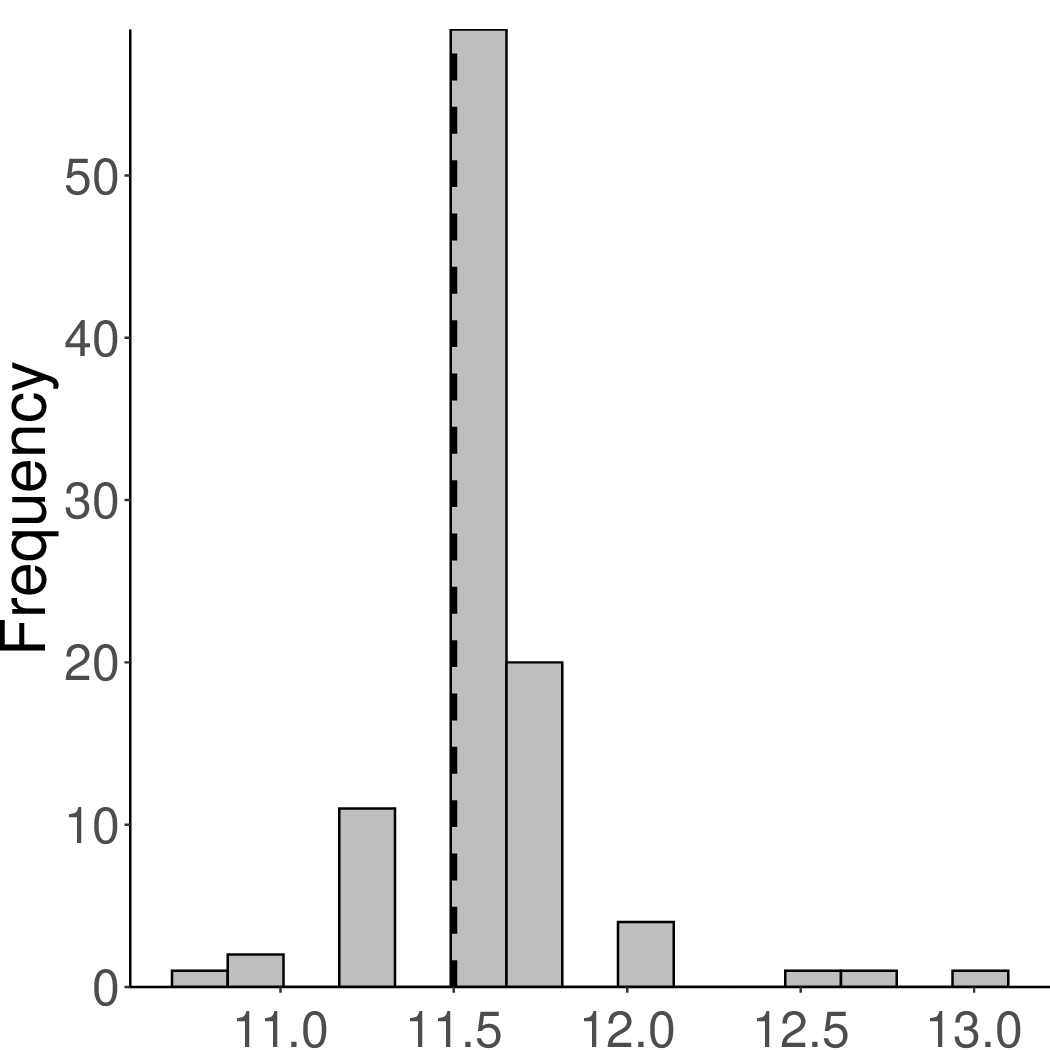}%
	\includegraphics[width=0.25\linewidth]{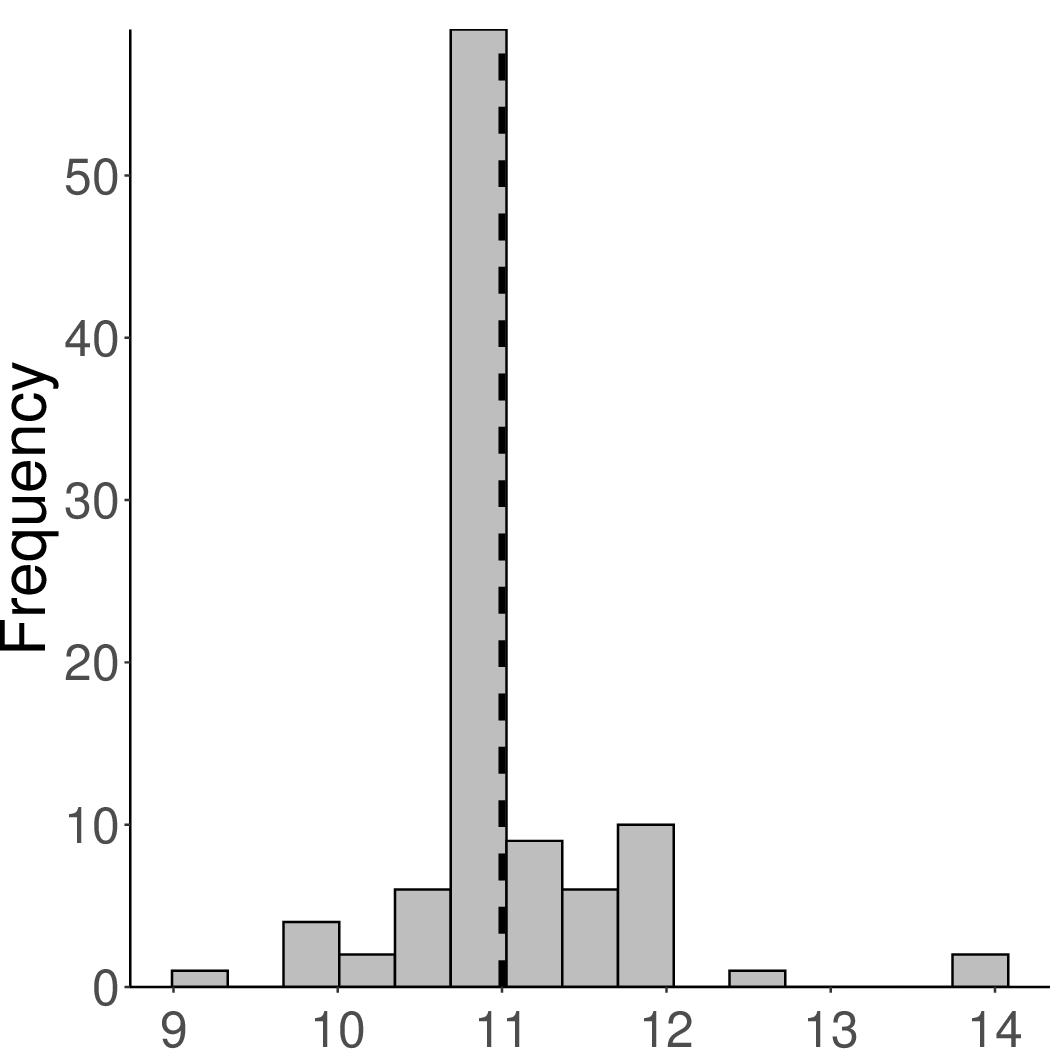}%
	\includegraphics[width=0.25\linewidth]{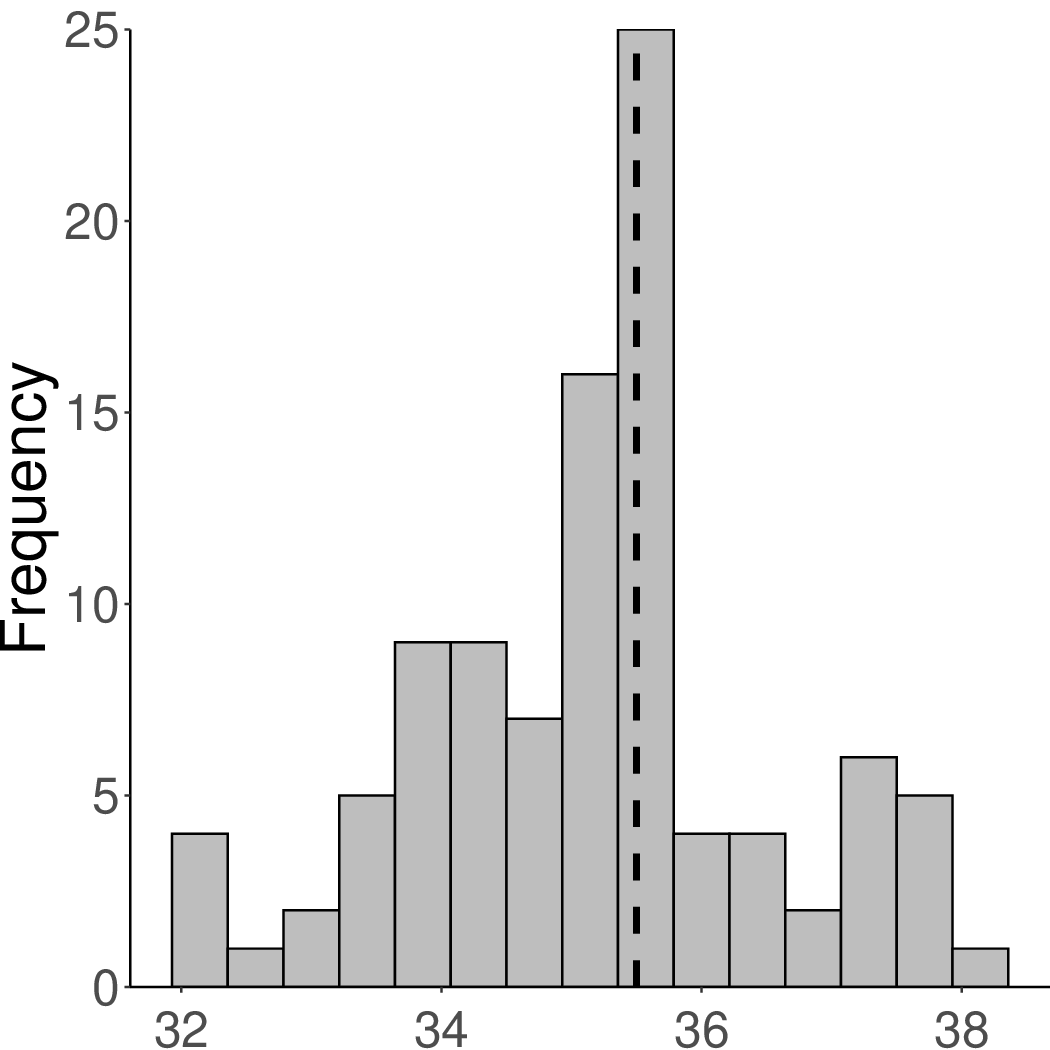}%
	\caption{
		Distribution of the parameter estimates obtained under the simulation study.
		The histograms are in reading order associated to ${\gamma_1}, {\gamma_2}, {h}, {r}_1, {t}_1, {r}_2, {t}_2$.
	}
	\label{f:simstud_hist}
\end{figure}

The results of the simulation study are presented in Table~\ref{t:simstud_parms} and Figure~\ref{f:simstud_hist}. Despite the high number of parameters, the estimates are distributed with most mass close to true parameter values and with a small standard deviation. Not surprisingly, the hard core parameter $h$ is an exception that is biased with a right skewed distribution, since the simulated point patterns always have to satisfy the hard core condition, whilst for a given simulated point pattern 
the estimate of $h$ is 
the shortest distance between any two points.

\bibliographystyle{apalike} 
\bibliography{references}

\begin{thebibliography}{}

\bibitem[Baddeley et~al., 2016]{BRT:2016}
Baddeley, A., Rubak, E., and Turner, R. (2016).
\newblock {\em Spatial Point Patterns: Methodology and Applications with R}.
\newblock Chapman \& Hall/CRC, New York.

\bibitem[Besag, 1975]{besag:75}
Besag, J. (1975).
\newblock Statistical analysis of non-lattice data.
\newblock {\em The Statistician}, 24:179--195.

\bibitem[Biscio and Lavancier, 2016]{Biscio:etal:2016}
Biscio, C. A.~N. and Lavancier, F. (2016).
\newblock Quantifying repulsiveness of determinantal point processes.
\newblock {\em Bernoulli}, 22:2001--2028.

\bibitem[Buxhoeveden et~al., 2002]{Buxhoeveden:etal:2002}
Buxhoeveden, D., Fobbs, A., Roy, E., and Casanova, M. (2002).
\newblock Quantitative comparison of radial cell columns in children with
  {D}own’s syndrome and controls.
\newblock {\em Journal of Intellectual Disability Research}, 46:76--81.

\bibitem[Buxhoeveden and Casanova, 2002]{Buxhoeveden:Casanova:2002}
Buxhoeveden, D.~P. and Casanova, M.~F. (2002).
\newblock The minicolumn hypothesis in neuroscience.
\newblock {\em Brain}, 125:935--951.

\bibitem[Casanova, 2007]{Casanova:2007}
Casanova, M.~F. (2007).
\newblock Schizophrenia seen as a deficit in the modulation of cortical
  minicolumns by monoaminergic systems.
\newblock {\em International Review of Psychiatry}, 19:361--372.

\bibitem[Casanova et~al., 2006]{Casanova:etal:2006}
Casanova, M.~F., {van Kooten}, I. A.~J., Switala, A.~E., {van Engeland}, H.,
  Heinsen, H., Steinbusch, H. W.~M., Hof, P.~R., Trippe, J., Stone, J., and
  Schmitz, C. (2006).
\newblock Minicolumnar abnormalities in autism.
\newblock {\em Acta Neuropathologica (Berl)}, 112:287--303.

\bibitem[Chance et~al., 2011]{Chance:etal:2011}
Chance, S.~A., Clover, L., Cousijn, H., Currah, L., Pettingill, R., and Esiri,
  M.~M. (2011).
\newblock Microanatomical correlates of cognitive ability and decline: {N}ormal
  ageing, {MCI}, and {A}lzheimer’s disease.
\newblock {\em Cerebral Cortex}, 21:1870–1878.

\bibitem[Daley and Vere-Jones, 2003]{DVJ:2003}
Daley, D.~J. and Vere-Jones, D. (2003).
\newblock {\em An Introduction to the Theory of Point Processes. Volume I:
  Elementary Theory and Methods}.
\newblock Springer-Verlag, New York, second edition.

\bibitem[Diggle, 2014]{D:2014}
Diggle, P.~J. (2014).
\newblock {\em Statistical Analysis of Spatial and Spatio-temporal Point
  Patterns}.
\newblock Chapman \& Hall/CRC, Boca Raton, Florida.

\bibitem[Diggle and Gratton, 1984]{Diggle:et:al:1984}
Diggle, P.~J. and Gratton, R.~J. (1984).
\newblock Monte {C}arlo methods of inference for implicit statistical models
  (with discussion).
\newblock {\em Journal of the Royal Statistical Society: Series B (Statistical
  Methodology)}, 46:193--227.

\bibitem[Esiri and Chance, 2006]{Esiri:Chance:2006}
Esiri, M.~M. and Chance, S.~A. (2006).
\newblock Vulnerability to {A}lzheimer's pathology in neocortex: the roles of
  plasticity and columnar organization.
\newblock {\em Journal of Alzheimer's Disease}, 9:79--89.

\bibitem[Guan, 2006]{Guan:2006}
Guan, Y. (2006).
\newblock A composite likelihood approach in fitting spatial point process
  models.
\newblock {\em Journal of the Americal Statistical Association},
  101:1502--1512.

\bibitem[Guan, 2009]{Guan:2009}
Guan, Y. (2009).
\newblock A minimum contrast estimation procedure for estimating the
  second-order parameters of inhomogeneous spatial point processes.
\newblock {\em Statistics and Its Interface}, 2:91--99.

\bibitem[Lavancier et~al., 2015]{Lavancier:etal:2015}
Lavancier, F., M{\o}ller, J., and Rubak, E. (2015).
\newblock Determinantal point process models and statistical inference.
\newblock {\em Journal of the Royal Statistical Society: Series B (Statistical
  Methodology)}, 77:853--877.

\bibitem[Lavancier et~al., 2018]{Lavancier:etal:2018}
Lavancier, F., Poinas, A., and Waagepetersen, R.~P. (2018).
\newblock Adaptive estimating function inference for non-stationary
  determinantal point processes.
\newblock Available on arXiv:1806.06231.

\bibitem[M{\o}ller et~al., 2019]{Moller:etal:2019}
M{\o}ller, J., Christensen, H.~S., Cuevas-Pacheco, F., and Christoffersen,
  A.~D. (2019).
\newblock Structured space-sphere point processes and {$K$}-functions.
\newblock {\em Methodology and Computing in Applied Probability}.
\newblock Available at {https://doi.org/10.1007/s11009-019-09712-w}.

\bibitem[M{\o}ller and Christoffersen, 2018]{Moller:Christoffersen:2018}
M{\o}ller, J. and Christoffersen, A.~D. (2018).
\newblock Pair correlation functions and limiting distributions of iterated
  cluster point processes.
\newblock {\em Journal of Applied Probability}, 55:789--809.

\bibitem[M{\o}ller et~al., 2016]{Moller:etal:2016}
M{\o}ller, J., Safavimanesh, F., and Rasmussen, J.~G. (2016).
\newblock The cylindrical {$K$}-function and {P}oisson line cluster point
  process.
\newblock {\em Biometrika}, 103:937--954.

\bibitem[M{\o}ller and Torrisi, 2005]{Moller:Torrisi:2005}
M{\o}ller, J. and Torrisi, G.~L. (2005).
\newblock Generalised shot noise {C}ox processes.
\newblock {\em Advances in Applied Probability}, 37:48--74.

\bibitem[M{\o}ller and Waagepetersen, 2004]{MW:2004}
M{\o}ller, J. and Waagepetersen, R.~P. (2004).
\newblock {\em Statistical Inference and Simulation for Spatial Point
  Processes}.
\newblock Chapman \& Hall/CRC, Boca Raton, Florida.

\bibitem[Mrkvi{\v c}ka et~al., 2017]{Mrkvicka:etal:2017}
Mrkvi{\v c}ka, T., Myllym{\"a}ki, M., and Hahn, U. (2017).
\newblock Multiple {M}onte {C}arlo testing, with applications in spatial point
  processes.
\newblock {\em Statistics and Computing}, 27:1239--1255.

\bibitem[Mrkvi{\v c}ka et~al., 2018]{Mrkvicka:etal:2018}
Mrkvi{\v c}ka, T., Myllym{\"a}ki, M., J\'{i}lek, M., and Hahn, U. (2018).
\newblock A one-way {ANOVA} test for functional data with graphical
  interpretation.
\newblock Available on arXiv:1612.03608.

\bibitem[Mrkvi{\v c}ka et~al., 2016]{Mrkvicka:etal:2016}
Mrkvi{\v c}ka, T., Soubeyrand, S., Myllym{\"a}ki, M., Grabarnik, P., and Hahn,
  U. (2016).
\newblock Monte {C}arlo testing in spatial statistics, with applications to
  spatial residuals.
\newblock {\em Spatial Statistics}, 18:40--53.

\bibitem[Myllym{\"a}ki et~al., 2017]{Myllymaki:etal:2017}
Myllym{\"a}ki, M., Mrkvi{\v c}ka, T., Grabarnik, P., Seijo, H., and Hahn, U.
  (2017).
\newblock Global envelope tests for spatial processes.
\newblock {\em Journal of the Royal Statistical Society: Series B (Statistical
  Methodology)}, 79:381--404.

\bibitem[Ogata and Katsura, 1991]{Ogata:etal:1991}
Ogata, Y. and Katsura, K. (1991).
\newblock Maximum likelihood estimates of the fractal dimension for random
  spatial patterns.
\newblock {\em Biometrika}, 78:463--474.

\bibitem[Proke\v{s}ov\'{a} et~al., 2016]{Prokesova:etal:2016}
Proke\v{s}ov\'{a}, M., Dvo\v{r}\'{a}k, J., and Jensen, E. B.~V. (2016).
\newblock Two-step estimation procedures for inhomogeneous shot-noise {C}ox
  processes.
\newblock {\em Annals of the Institute of Statistical Mathematics}, 69:1--30.

\bibitem[Rafati et~al., 2016]{Rafati:etal:2016}
Rafati, A.~H., Safavimanesh, F., Dorph-Petersen, K.-A., Rasmussen, J.~G.,
  M{\o}ller, J., and Nyengaard, J.~R. (2016).
\newblock Detection and spatial characterization of minicolumnarity in the
  human cerebral cortex.
\newblock {\em Journal of Microscopy}, 261:115--126.

\bibitem[Ripley, 1976]{Ripley:1976}
Ripley, B.~D. (1976).
\newblock The second-order analysis of stationary point processes.
\newblock {\em Journal of Applied Probability}, 13:255--266.

\bibitem[Thomas, 1949]{Thomas:1949}
Thomas, M. (1949).
\newblock A generalization of {P}oisson's binomial limit for use in ecology.
\newblock {\em Biometrica}, 36:18--25.

\bibitem[{Van Lieshout} and Baddeley, 1996]{Lieshout:Baddeley:1996}
{Van Lieshout}, M. N.~M. and Baddeley, A.~J. (1996).
\newblock A nonparametric measure of spatial interaction in point pattersn.
\newblock {\em Statistica Neerlandica}, 50:344--361.

\bibitem[Waagepetersen, 2007]{Waagepetersen:2007}
Waagepetersen, R.~P. (2007).
\newblock An estimating function approach to inference for inhomogeneous
  {N}eyman-{S}cott processes.
\newblock {\em Biometrics}, 63:252--258.

\end{thebibliography}
\end{document}